\documentclass[iop]{emulateapj-rtx4}
\usepackage{bm}
\usepackage{threeparttable}
\usepackage{CJK}
\def\msun{M_\odot}
\def\mbh{M_{\rm{BH}}}

\renewcommand\bv{{\mbox{\boldmath $v$}}}
\newcommand\bb{{\mbox{\boldmath $B$}}}

\def\<{\,\langle\langle}
\def\>{\,\rangle\rangle}

\begin{document}
\begin{CJK*}{UTF8}{gbsn}

\shortauthors{Y.-F. Jiang et al.}
\author{Yan-Fei Jiang(姜燕飞)\altaffilmark{1,4}, Shane W. Davis\altaffilmark{2} \& James M. Stone\altaffilmark{3} }
%\author{Yan-Fei Jiang\altaffilmark{1,4}, Shane W. Davis\altaffilmark{2} \& James M. Stone\altaffilmark{3}}
\affil{$^1$Harvard-Smithsonian Center for Astrophysics, 60 Garden Street, Cambridge, MA 02138, USA} 
\affil{$^2$Department of Astronomy, University of Virginia, P.O. Box 400325, Charlottesville, VA 22904-4325, USA}
\affil{$^3$Department of Astrophysical Sciences, Princeton University, Princeton, NJ 08544, USA} 
\altaffiltext{4}{Einstein Fellow}

\title{Iron Opacity Bump Changes the Stability and Structure of  Accretion Disks in Active Galactic Nuclei}

\begin{abstract}
Accretion disks around supermassive black holes have regions 
where the Rosseland mean opacity can be much larger than the electron 
scattering opacity primarily due to the large number of bound-bound transitions
 in iron. We study the effects of this iron opacity ``bump'' on the thermal stability and vertical structure of radiation pressure 
dominated accretion disks, utilizing three dimensional radiation magneto-hydrodynamic
simulations in the local shearing box approximation. The simulations self-consistently calculate the heating due to 
MHD turbulence caused by magneto-rotational instability and radiative cooling 
by using the radiative transfer module  based on a variable Eddington tensor in {\sc Athena}. 
For a $5\times 10^8$ solar mass black hole with $\sim 3\%$ 
of the Eddington luminosity, a model including the iron opacity bump 
maintains its structure for more than $10$ thermal times without showing significant signs of thermal runaway. 
In contrast, if only electron scattering and free-free opacity are included as in 
the standard thin disk model, the disk collapses on the thermal time scale. 
The difference is caused by a combination of 1) an anti-correlation between the total optical depth 
and the temperature, and 2) enhanced vertical advective energy transport. 
These results suggest that the iron opacity bump may have a strong impact on the stability and structure 
of AGN accretion disks, and may contribute to a dependence of 
AGN properties on metallicity. 
Since this opacity is relevant primarily in UV emitting regions of the flow, it may help to explain discrepancies between
observation and theory that are unique to AGNs.

\end{abstract}


\maketitle

\section{Introduction}
\label{sec:introduction}
The radiation from active galactic nuclei (AGNs) is widely believed 
to be produced from the gas accreted by supermassive black 
holes through their accretion disks \citep[e.g.,][]{Malkan1983}. The standard $\alpha$ disk model \citep[][]{ShakuraSunyaev1973} 
is commonly used to model the structures of accretion disks in most AGNs, 
mainly because the temperature predicted by the $\alpha$ disk model is roughly 
consistent with the big blue bump in AGN spectra \citep[][]{Elvisetal1994,KoratkarBlaes1999}. 
However, there are discrepancies when detailed predictions of the $\alpha$ disk model are 
compared with AGN observations \citep[see e.g.][]{KoratkarBlaes1999}.  

From the theoretical point of view, the main puzzle is the outcome of the inflow (``viscous'') and thermal instabilities
in $\alpha$ disks that are radiation pressure dominated,
which should be the case for the inner region of AGN disks \citep[][]{LightmanEardley1974,ShakuraSunyaev1976}. The thermal instability,
which should grow faster than the inflow instability,
exists because the cooling rate $Q^-$ in this regime is proportional to the midplane 
pressure $P_{z,0}$ while the heating rate $Q^+\propto P_{z,0}^2$ \citep[][]{Piran1978}. When the pressure 
is perturbed to be smaller (larger) than equilibrium value, $Q^+$ decreases (increases) faster compared with the change of 
$Q^-$ and the disk will continue to cool down (heat up) and never return to the original equilibrium state.  The existence of the
radiation dominated thermal instability 
has been questioned \citep[e.g.][]{StellaRosner1984,Merloni2003}.  The first 3D local shearing box simulations with
turbulence driven by the magneto-rotational instability (MRI, \citealt[][]{BalbusHawley1991}) and optically thick radiative cooling
seemed consistent with thermal stability \citep[][]{Turner2004,Hiroseetal2009}, but more recent simulations performed in larger domains
and using more sophisticated radiative transfer methods  have found that the radiation dominated 
accretion disks still show signs of thermal runaway after a few thermal time scales \citep{Jiangetal2013c}. 
However, no clear evidence of rapid variability that might be consistent
with the thermal instability has yet been observed in most AGNs.  Although there has been speculations that recently discovered ``change-look'' AGNs 
\citep[][]{LaMassaetal2015,Ruanetal2015} may be caused by these instabilities, it is difficult to understand why the instabilities would
uniquely manifest themselves only in this specific subset of AGNs.

Observationally, many properties of AGNs cannot be explained with the standard $\alpha$ disk model \citep[][]{KoratkarBlaes1999}, 
in spite of the issue of thermal instability. The Spectrum Energy Distribution (SED) of most AGNs shows a turnover always 
around $1000$ {\r A}, almost independent of the black hole mass \citep[][and references therein]{Zhengetal1997,Bonningetal2007,Davisetal2007,LaorDavis2014}. 
The predicted Lyman edge is also not observed \citep[][]{Shulletal2012}. The inferred size of the AGN accretion disks based on 
micro-lensing measurements \citep[e.g.][]{Morganetal2010,Blackburneetal2011} or the lags between the variations in different continuous 
bands \citep[][]{Edelsonetal2015} is systematically larger than the half-light radius of the $\alpha$ disk model. A soft X-ray excess is 
also ubiquitously observed in bright AGNs \citep[][]{Crummyetal2006,Doneetal2012,Done2014}, but is not easily explained by the $\alpha$ disk model
if it results from continuum emission.
 
The order-of-magnitude temperature and density in the AGN accretion disks can actually be roughly estimated based on a few assumptions 
independent of any accretion disk model. If we assume the accreted gas is optically thick with typical Eddington 
luminosity $L_{\rm Edd}=1.5\times 10^{46} \mbh/(10^8\msun)$ erg/s, 
typical emission size to be the Schwarzschild radius $r_s=3.0\times 10^{13} \mbh/(10^8\msun)$ cm,  then the effective 
temperature is $T\sim3.9\times 10^5\left(\mbh/10^8\msun\right)^{-1/4}$ K. Density of the gas depends on the assumed 
inflow velocity, which is likely subsonic with respect to the radiation sound speed \citep[][]{Pringle1981}. As an order-of-magnitude estimate, we take it to 
be the gas isothermal sound speed with the effective temperature estimated above. With the typical mass accretion rate 
$\dot{M}_{\rm Edd}=10L_{Edd}/c^2=1.6\times 10^{26}\mbh/(10^8\msun)$ g/s and inflow radius $r_s$, the typical density is 
$\rho\sim 2\times 10^{-9}$ g/cm$^3$ for the $10^8\msun$ black hole. The density and temperature in the $\alpha$ disk model 
are consistent with this simple estimate. However, this temperature and density regimes are very similar as in the envelope of 
massive stars \citep[][]{Paxtonetal2011,Jiangetal2015}. With non-negligible metallicity, the OPAL opacity project \citep[][]{iglesias96} 
shows that the Rosseland mean opacity can be significantly enhanced compared with the electron scattering value 
due to metals. In particular, the irons produce the well-known opacity bump around the 
temperature $1.8\times 10^5$ K. The opacity drops rapidly with increasing or decreasing temperature and it only depends on 
the density weakly (Figure 2 of \citealt{Jiangetal2015}).  As significant metals are indeed observed  in the broad line 
regions of AGNs \citep{HamannFerland1993,Dhandaetal2007}, the iron opacity bump should exist in the AGN disks. However, 
the standard $\alpha$ disk model only includes the electron scattering and free-free opacities. 
Following \cite{Jiangetal2013c}, we will calculate the turbulence from MRI without any 
$\alpha$ assumption and radiative transfer including the iron opacity bump self-consistently based on 3D radiation 
MHD simulations. Because optical depth is critical 
for the radiative cooling of the disk, we will study  how the thermal stability and structures of the AGN disks will be affected by 
the iron opacity bump. Note that for accretion disks in X-ray binaries around stellar mass 
black holes, the disk midplane temperature in the inner region is too hot for iron opacity to play an important role, but it
may play a role in the outer disk dynamics. 
Another example where opacity effects significantly change the dynamics of the disk is the hydrogen ionization instability
in accretion disks around white dwarfs \citep[][]{Lasota2001,Hiroseetal2014}.

This paper is organized as follows.  We describe how we setup the simulations in 
\S~\ref{sec:setup}, and the initial and boundary conditions we use
in \S~\ref{sec:INIBDcondition}.  Our primary results are
described in \S~\ref{sec:result}, while \S~\ref{sec:discussion}
compares the simulation results with the $\alpha$ disk model and 
discusses the implications for AGN observations.

\section{Simulation Setup}
\label{sec:setup}
We solve the same set of radiation MHD equations under the local shearing box 
approximation as equation (2) of \cite{Jiangetal2013c} given by. 
The simulation box is located at a fiducial radius $r_0=20$ Schwarzschild radii from a $\mbh=5\times 10^8 M_\odot$
black hole. Keplerian rotation is assumed for the background flow with the orbital frequency 
$\Omega=1.60\times 10^{-6}$ s$^{-1}$ and shear parameter $q=-d\ln\Omega/d\ln r=3/2$. 
We solve these equations with the Godunov radiation MHD code  as
described and tested in \cite{Jiangetal2012} and \cite{Davisetal2012},
with the improvements given by \cite{Jiangetal2013b}. Under the shearing box 
approximation, heating is generated by the dissipation from MRI turbulence with 
the energy ultimately coming from work done by the shearing period boundary condition, while 
cooling is dominated by the photons leaving from the top and bottom of the simulation box. 

For the opacities describing the interactions between the 
radiation and gas, we calculate the total Rosseland mean opacity $\kappa_t$ based on the 
opacity table from the Modules for Experiments in Stellar Evolution (MESA, 
\citealt[][]{Paxtonetal2011}) for solar metallicity in order to capture the dependencies 
of the opacity on temperature and density correctly. This opacity table is originally from 
the OPAL opacity project \citep[][]{iglesias96} and has been successfully used to study 
the stability and structures of massive star envelopes \citep[][]{Jiangetal2015}. The opacity 
as a function of temperature and density is shown in Figure 2 of \cite{Jiangetal2015}, which 
also covers the relevant parameter space for AGN disks. The most important feature 
is the opacity bump caused by a large number of bound-bound transitions, most due to iron
atoms, for temperatures around $1.8\times 10^5$ K.  This 
can be a factor of $\sim 4$ larger than the electron scattering opacity for solar metallicity. 
In order to split this Rosseland mean opacity 
into scattering and absorption terms (as only the absorption opacity enters the energy exchange 
terms in the radiation moment equations), we simply adopt the electron 
scattering opacity $\kappa_{\rm es}=0.34$ cm$^2$ g$^{-1}$ and subtract this from the Rosseland 
mean opacity. The Planck mean absorption opacity is taken to be the 
same as Rosseland mean absorption opacity for simplicity, although it is likely to be an underestimate 
since the Rosseland mean tends to more heavily weight frequencies with low opacity. For comparison, we also perform simulations 
with the ``standard'' thin disk model opacity, 
which only includes the electron scattering opacity 
$\kappa_{\rm es}$, Plank-mean free-free
absorption opacity $\kappa_{\rm aP}=3.7\times10^{53}\left(\rho^9/E_g^7\right)^{1/2}$
cm$^2$ g$^{-1}$ and Rosseland-mean free-free absorption opacity
$\kappa_{\rm aF}=1.0\times10^{52}\left(\rho^9/E_g^7\right)^{1/2}$ cm$^2$ g$^{-1}$. 
Here $E_g=P_g/(\gamma-1)$ is the gas internal energy with gas pressure $P_g$, density $\rho$ 
and adiabatic index $\gamma=5/3$. This combination of opacities has been used in most previous simulations,
which focussed on hotter disks more appropriate to $\sim 10 M_\odot$ black hole X-ray binaries \citep{Hiroseetal2009,Jiangetal2013c}.

\section{Initial and Boundary Conditions}
\label{sec:INIBDcondition}
We construct the initial vertical profiles of the disk based on hydrostatic 
equilibrium and diffusion equation as in \cite{Hiroseetal2009} and 
\cite{Jiangetal2013c} but with 
the the iron opacity bump included self-consistently. We assume the dissipation 
profile $dF_{r,z}/dz\propto \rho\kappa_t/\tau^{0.5}$, where $\tau$ 
is the optical depth from the nearest surface of the disk and 
$F_{r,z}$ is the vertical component of the radiation flux. If  
the total optical depth from the disk midplane to the surface is $\tau_0$, 
by symmetry, the radiation flux as a function of $\tau$ within the photosphere is 
$F_{r,z}=F_{\rm max}\left(\tau_0^{0.5}-\tau^{0.5}\right)/\left(\tau_0^{0.5}-1\right)$. 
Here we choose the initial maximum radiation flux $F_{\rm max}=7.92\times 10^{11}$ erg s$^{-1}$ cm$^{-2}$, 
and $F_{r,z}$ is fixed to this value in the region where $\tau<1$. 
We choose the midplane temperature $2.4\times 10^5$ K and integrate vertically 
according to the diffusion equation $dE_r/d\tau=3 F_{r,z}/c$, where $c$ is the speed of light. 
The total optical depth $\tau_0$ is chosen such that at $\tau=1$, $E_r=\sqrt{3}cF_{r,z}$. 
Initially, gas temperature $T$ is set to 
be the same as the radiation temperature $T_r\equiv\left(E_r/a_r\right)^{0.25}$, where the radiation 
constant $a_r=7.57\times 10^{15}$ erg cm$^{-3}$ K$^{-4}$. The initial density profile is constructed based 
on the equation of hydrostatic equilibrium $dP/dz=\kappa_tF_{r,z}/c-\Omega^2z$ and $d\tau=-\rho \kappa_tdz$. 
The midplane density is adjusted to be $10^{-8}$ g cm$^{-3}$ such that the initial total 
optical depth is $\tau_0=9.5\times 10^5$. All the quantities above the photosphere are fixed to be the same values 
as at $\tau=1$. Notice that in the initial condition, only the surface density $\Sigma$ is the conserved quantity, while 
all the other quantities such as $\rho, T, \tau_0, F_{r,z}$ will adjust self-consistently during the simulation. 
The magnetic field is initialized in the same way as in \cite{Jiangetal2013c} with the initial ratio between gas pressure and 
magnetic pressure to be $12$ at $z=0$. The boundary conditions are also the same as in \cite{Jiangetal2013c}. For the 
short characteristic module we use to calculate the variable Eddington tensor (VET), we use $80$ angles per cell 
to capture the angular distribution of the radiation field. 
Sizes of the simulation box are all fixed to be $L_x=0.87H_s$, 
$L_y=3.48H_s$ and $L_z=6.96H_s$, where $H_s$ is the length unit listed in Table 
\ref{table:parameters}. The length unit is chosen based the total radiation flux $F_{\rm max}$ we get from the 
simulation (Table \ref{table:parameters}) as $H_s=\kappa_{\rm es}F_{\rm max}/(c\Omega^2)$. 
For the typical density $\rho_0$ and temperature $T_0$ given in table 
\ref{table:parameters}, it is related to the gas pressure scale height $H_g\equiv c_g/\Omega$ and 
radiation pressure scale height $H_r\equiv c_r/\Omega$ as $H_s=8.57H_g=2.25H_r$, where $c_g$ 
is the isothermal sound speed for temperature $T_0$ and radiation sound speed $c_r=\sqrt{a_rT_0^4/(3\rho_0)}$. 
We use $64\times 128\times 512$ grids for $x,y,z$ directions so that we have roughly $32$ grids per radiation 
pressure scale height $H_r$. Following the convention in {\sc Athena} \citep[][]{Stoneetal2008}, unit of the magnetic field is chosen so that 
magnetic permeability is one.

\section{Results}
\label{sec:result}

The initial evolutions of the disks are very similar for the cases with or without the 
iron opacity. The disk cools down slightly while MRI is still growing from the laminar initial 
condition during the first few orbits. However, once heating is generated by 
vigorous MHD turbulence from MRI,  the disk undergoes quite different 
evolution histories for different opacities. We label the run with iron opacity 
bump as {\sf OPALR20}, while three runs with just electron scattering and free-free opacities for 
comparison as {\sf ESR20a}, {\sf ESR20b} and {\sf ESR20c}. 

\begin{table}[htp]
\centering
\caption{Simulation Parameters of the run {\sf OPALR20}}
\begin{tabular}{cc}
\hline\hline
$\Omega$ / s$^{-1}$     & $1.60\times 10^{-6}$ \\
$r_0$ / cm		    &  $2.97\times 10^{15}$ \\
$\rho_0$/ g cm$^{-3}$  & $1.00\times 10^{-8}$ \\
$T_0$ / K			    & $2.00\times 10^5$ \\
$P_0$ / dyn cm$^{-2}$  & $2.77\times 10^5$ \\
$H_s$ / cm		    & $2.81\times 10^{13}$ \\
$\Sigma$ / g cm$^{-2}$ & $4.34\times 10^5$ \\
$\tau_0$			  & $2.31\times 10^5$ \\
$F_{\rm max}$/erg s$^{-1}$ cm$^{-2}$    & $6.36\times 10^{12}$ \\ 
\hline
\end{tabular}
\begin{tablenotes}
\item Note: The parameters $\Omega$, $r_0$ and $\Sigma$ 
are fixed for the simulation, while $\tau_0$ and $F_{\rm max}$ 
are time averaged properties between $60$ and $125$ orbits. 
The density $\rho_0$, pressure $P_0$, temperature $T_0$ and scale height $H_s$ 
are the fiducial units we use to describe the simulation. They are pretty close to 
the midplane density, temperature and characteristic disk scale height. 
\end{tablenotes}
\label{table:parameters}
\end{table}

\subsection{Simulation History}

\begin{figure}[htp]
\centering
\includegraphics[width=1.0\hsize]{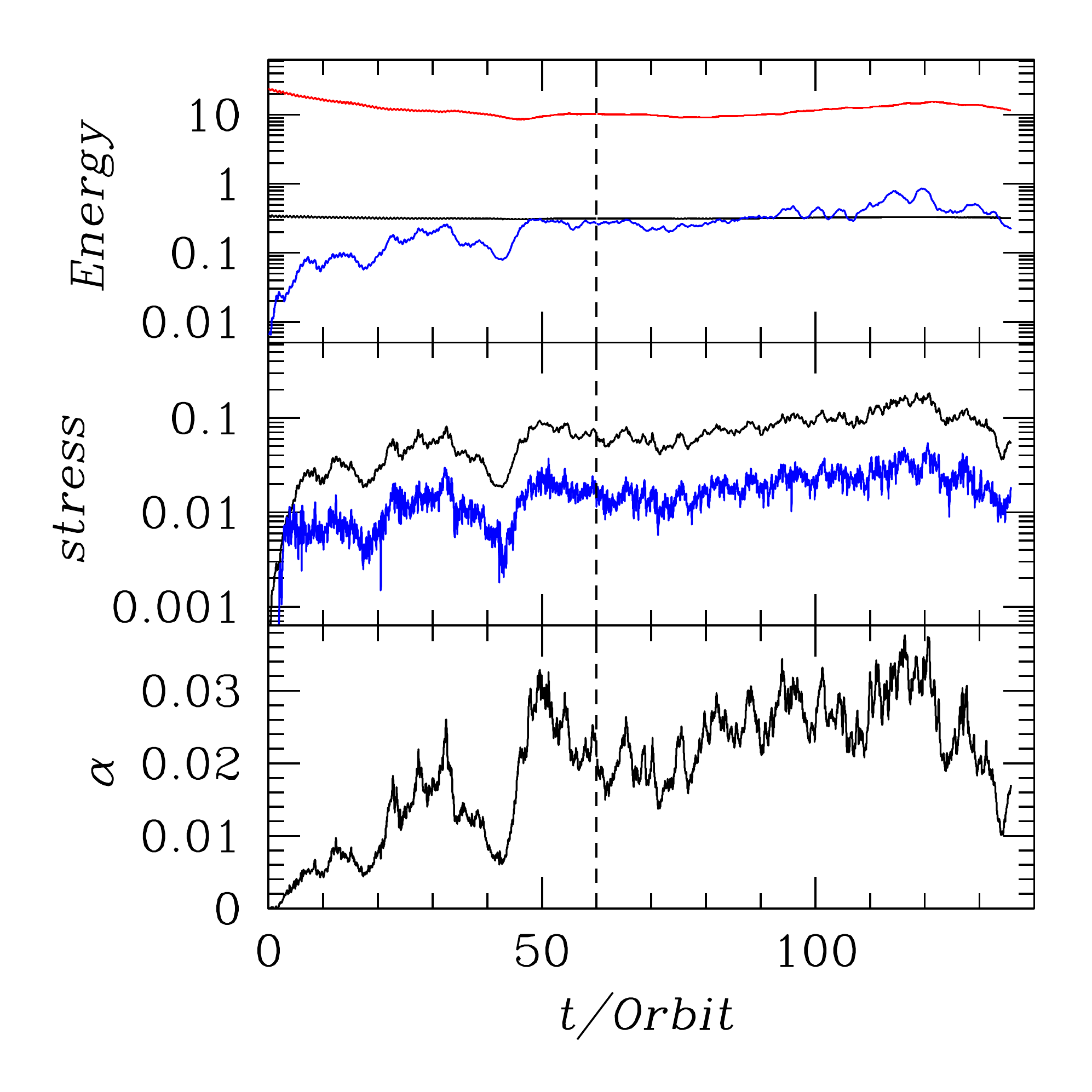}
\caption{Top: histories of the volume averaged radiation energy density 
$E_r$ (red line), gas internal energy $E_g$ (black line) and magnetic energy 
density $E_B$ (blue line) for the run {\sf OPALR20} with iron opacity bump. 
Middle: history of the volume averaged Maxwell stress $-B_xB_y$ (black line) 
and Reynolds stress $\rho v_x\delta v_y$ (blue line). Bottom: history of the $\alpha$ parameter, 
which is the ratio between the sum of the total Maxwell and Reynolds 
stress and the sum of the total radiation, gas and magnetic pressure. The vertical 
dashed line separates the first $60$ orbits when Eddington approximation is adopted 
and the time with VET turned on. Units for the energy density and stress are $P_0$ as given 
in Table \ref{table:parameters}.}
\label{OPALR20History}
\end{figure}

\begin{figure}[htp]
\centering
\includegraphics[width=1.0\hsize]{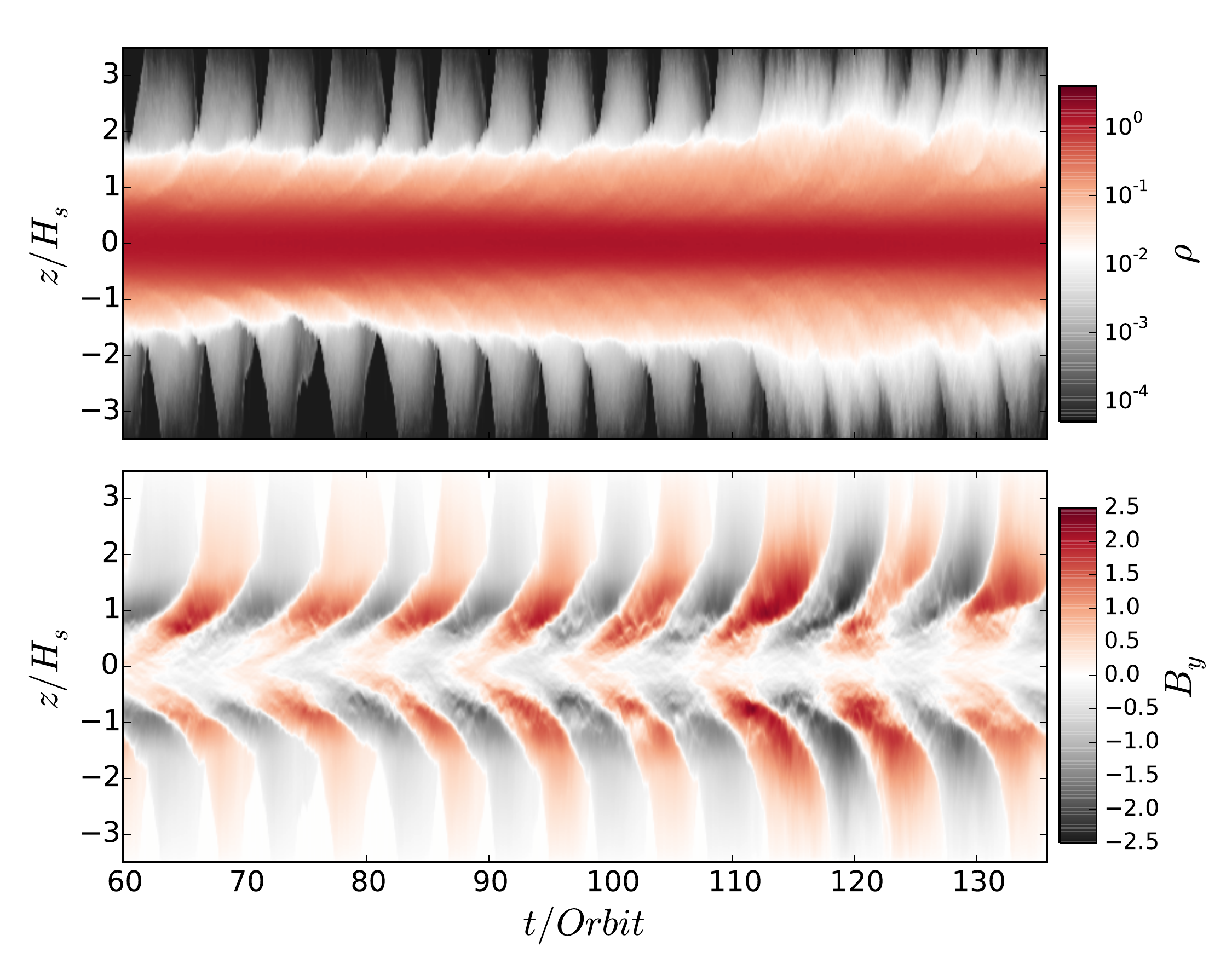}
\caption{Space-time diagram of the density $\rho$ (top panel, in unit of $\rho_0$) 
and azimuthal magnetic field $B_y$ (bottom panel, in unit of $\sqrt{2 P_0}$) for 
the run {\sf OPALR20}. }
\label{OPALR20STplot}
\end{figure}

\subsubsection{The Run {\sf OPALR20} with Iron Opacity Bump}
For {\sf OPALR20}, we first ran the simulation by setting 
the Eddington tensor ${\sf f}=1/3{\sf I}$ and disk lasted for $60$ orbits. Then the simulation 
is restarted with the short characteristic module turned on to calculate the VET 
self-consistently for another $75$ orbits. Histories of the volume averaged energy densities for the 
whole simulation duration are shown in the top panel of Figure \ref{OPALR20History}.
Although we have run the simulation for more than $10$ thermal time scales and  $E_r$ is more 
than $60$ times larger than $E_g$, the radiation energy 
density $E_r$, gas internal energy $E_g$ and magnetic energy density $E_b$ do not show 
any thermal runaway behavior as shown in \cite{Jiangetal2013c}.  
The radiation energy density $E_r$
varies only by a factor of $2$ while $E_g$ is almost a constant. Magnetic energy density $E_b$ shows 
a larger variation amplitude and it can change by a factor of $5$ after the initial $60$ orbits. 
Histories of the volume averaged Maxwell stress and Reynolds stress are shown in the middle 
panel of Figure \ref{OPALR20History}, while history of the equivalent $\alpha$ parameter is shown in the bottom 
panel of Figure \ref{OPALR20History}.  Here $\alpha$ is calculated as the ratio between the time and volume 
averaged stress and total pressure, which is $0.025$ after the first $60$ orbits. 
The average ratio between the total Maxwell stress and Reynolds stress from the MRI turbulence 
is $4.33$ while the average ratio between the total Maxwell stress and magnetic pressure is $0.25$. 
These statistical properties are consistent with previous local shearing box or global simulations of 
MRI turbulence, either with isothermal equation of state or self-consistent radiative transfer 
\citep[][]{Turneretal2003, Guanetal2009,Hawleyetal2011,Sorathiaetal2012,Jiangetal2013b}.

The space-time diagram of the density $\rho$ and toroidal magnetic field 
$B_y$ for this simulation after the first $60$ orbits are shown in Figure 
\ref{OPALR20STplot}, where the well-known butterfly diagram is clearly observed. 
The turbulent magnetic field generated by MRI peaks around $z\approx \pm H_s$.  The butterfly pattern
can be attributed to regions of strong $B_y$ buoyantly rising away from the midplane. 
This pattern of $B_y$ reverses roughly every $10$ orbits. The buoyantly 
rising magnetic field provides enhanced pressure support and causes the density scale height near the surface
to rise and fall following the same pattern. 
These buoyant motions also affect the energy transport inside the disk, and are discuss further
in section \ref{VerticalStructure}. 

\begin{figure}[htp]
\centering
\includegraphics[width=1.0\hsize]{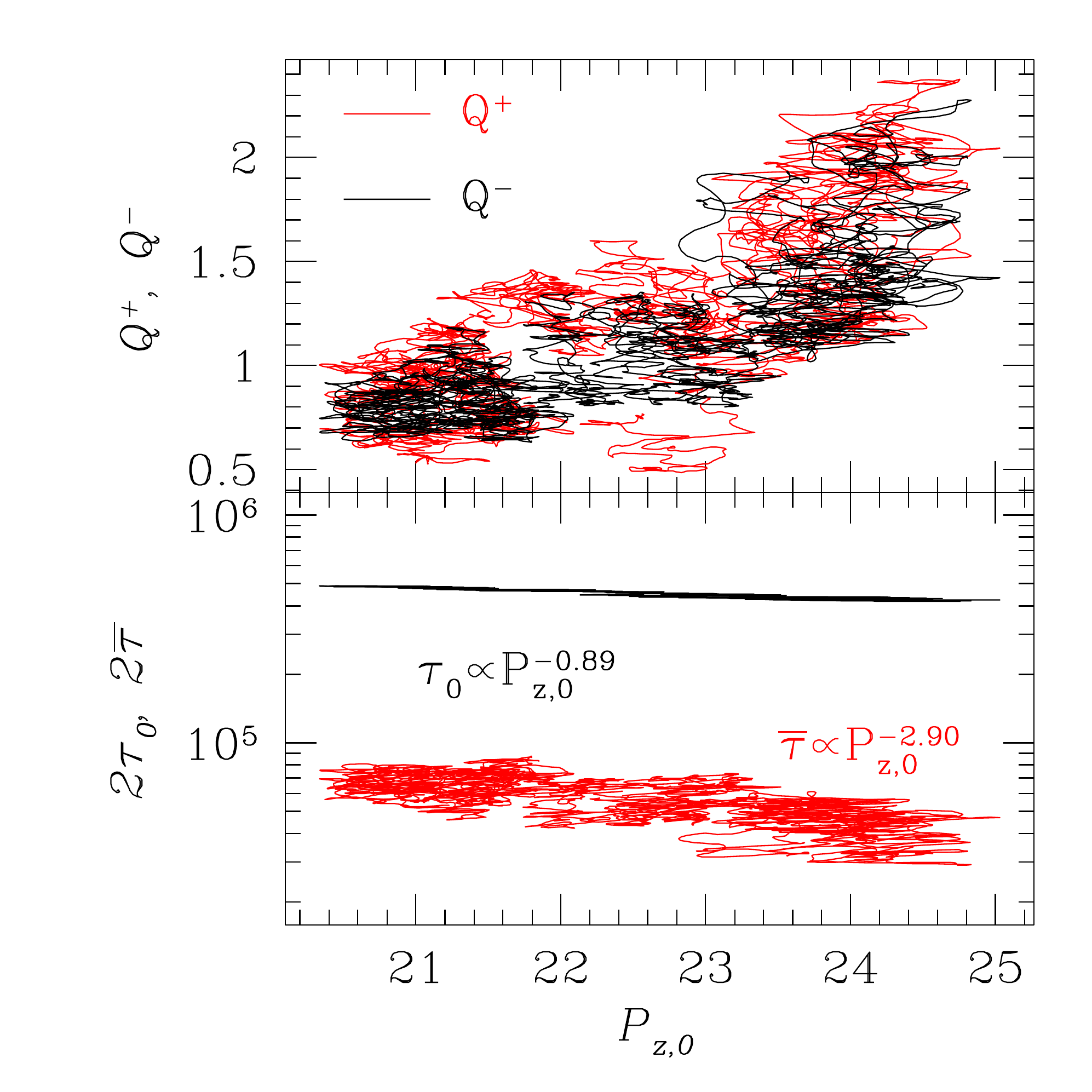}
\caption{Top: change of the heating $Q^+$ and cooling $Q^-$ rates 
per unit area as a function of the midplane total pressure $P_{z,0}$ 
for the run {\sf OPALR20}. Bottom: change of the total optical depth $\tau_0$ and 
flux weighted optical depth $\bar{\tau}$ 
as a function of $P_{z,0}$. The midplane pressure is in unit of $P_0$ 
while the units for $Q^+$ and $Q^-$ are $P_0H_s\Omega$. }
\label{HeatingCooling}
\end{figure}

To confirm that the disk is in roughly thermal equilibrium, we compare the 
total heating $Q^+$ and cooling $Q^-$ rates according to equations (2) and (3) in \cite{Jiangetal2013c}, 
which are shown in the top panel of Figure \ref{HeatingCooling}. Indeed, $Q^+$ 
and $Q^-$ track each other very well and midplane pressure only varies in a very small dynamic range. 
This is quite different from Figure 2 of \cite{Jiangetal2013c}, where $Q+$ and $Q^-$ diverge from each other when the disk 
undergoes thermal runaway. Because the disk is in thermal equilibrium, we also cannot measure the 
dependence of $Q^+$ and $Q^-$ on $P_{z,0}$ as we did in \cite{Jiangetal2013c}.

\subsubsection{Three Runs Without the Iron Opacity Bump}
To confirm that the different behaviors we see between the run {\sf OPALR20} and the simulations shown 
in \cite{Jiangetal2013c} are indeed caused by the iron opacity bump, we have done three comparison runs  
by only including the electron scattering and free-free opacities as in \cite{Jiangetal2013c}. For the run {\sf ESR20a}, 
we use the same surface density as in {\sf OPALR20}.  Because electron scattering opacity is smaller than the 
iron opacity, the total optical depth $\tau_0$ is only $34\%$ of the value in {\sf OPALR20}. For the run {\sf ESR20b}, 
we increase the surface density by a factor of $2$ so that the total optical depth is closer to the value in {\sf OPALR20}. 
We do not increase the surface density to match the $\tau_0$ in {\sf OPALR20}, because then the 
surface density is larger than the maximum surface density allowed by the thin disk model with $\alpha=0.02-0.03$. 
Initial conditions for the two runs are constructed in the same way as described in Section \ref{sec:INIBDcondition}. 
To test the effect of the initial condition, for the run {\sf ESR20c}, we restart the simulation 
{\sf OPALR20} at $60$ orbit by changing the opacity to be electron scattering and free-free opacities 
while keeping all the other quantities unchanged. In this way, {\sf ESR20c} 
has exactly the same turbulence as {\sf OPALR20} to start with. 
All the other parameters of the three runs, such as box size and resolution, are the same as the run {\sf OPALR20}. 

\begin{figure}[htp]
\centering
\includegraphics[width=1.0\hsize]{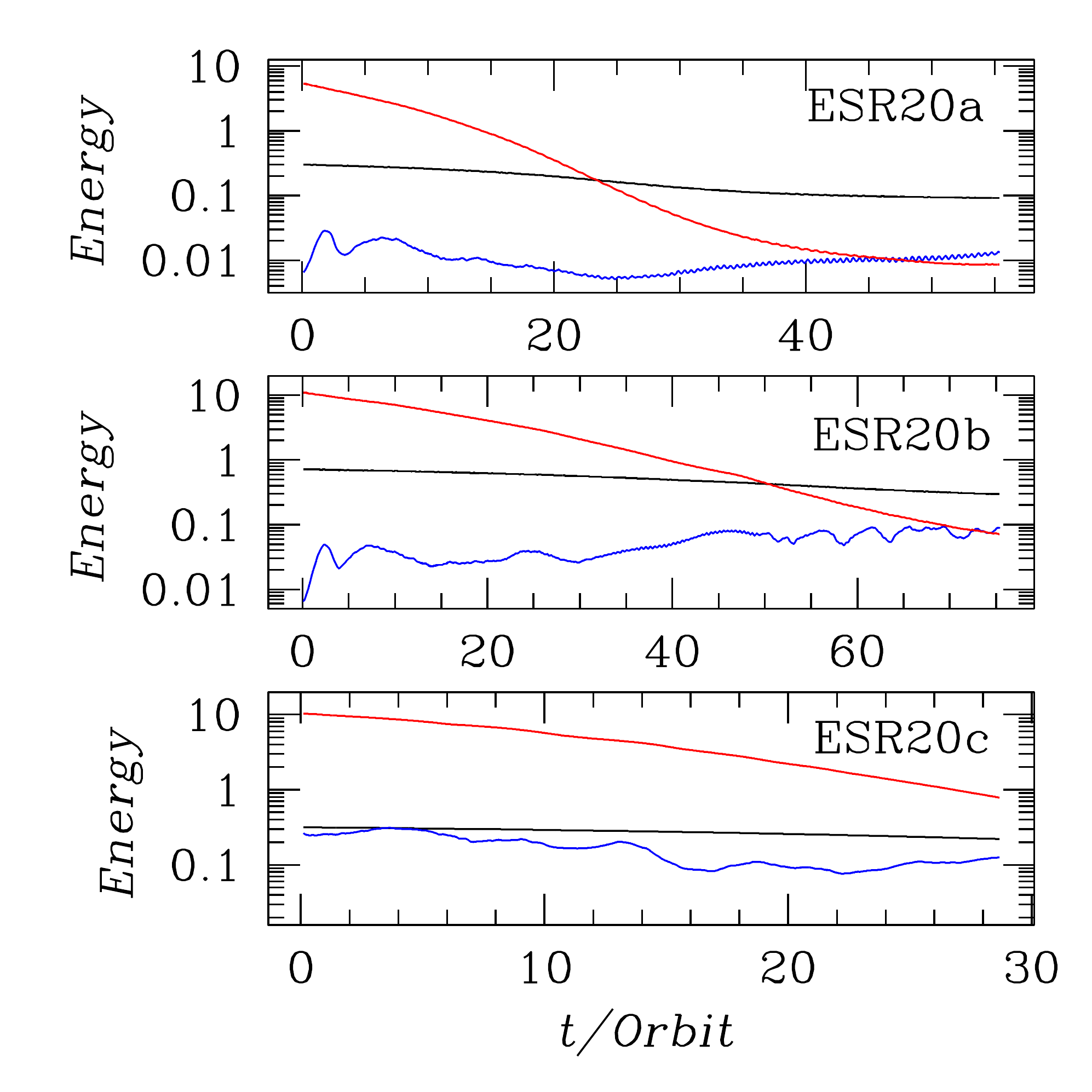}
\caption{Histories of the volume averaged energy densities $E_r$ (red lines), $E_g$ (black lines) 
and $E_B$ (blue lines). From the top to bottom, they are for the three runs {\sf ESR20a}, {\sf ESR20b} 
and {\sf ESR20c}, which only include the electron scattering and free-free opacities. Units of the energy 
densities are $P_0$. }
\label{CompareEnergyHist}
\end{figure}

\begin{figure}[htp]
\centering
\includegraphics[width=1.0\hsize]{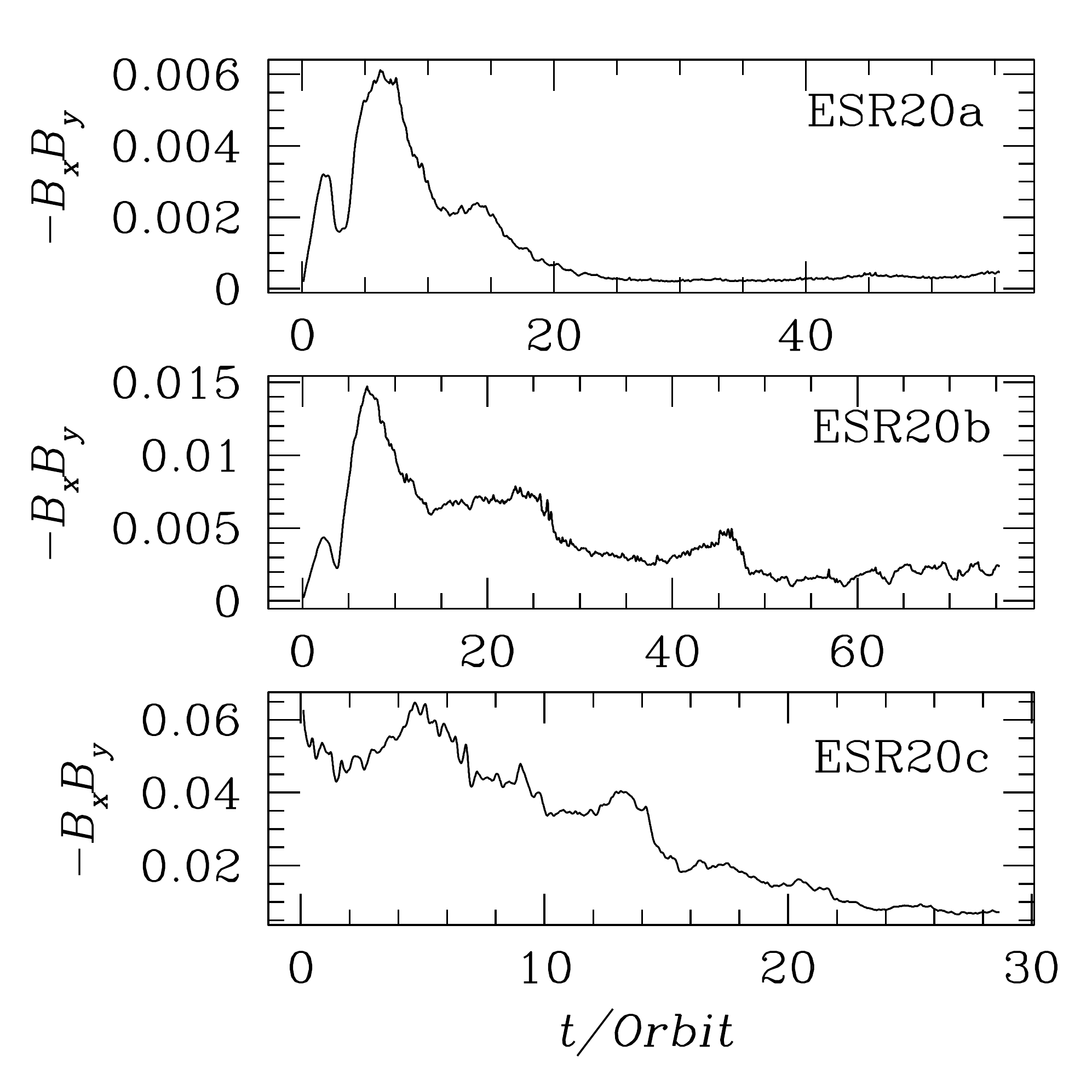}
\caption{Histories of the volume averaged Maxwell stress for the three runs 
{\sf ESR20a}, {\sf ESR20b} and {\sf ESR20c}. Units of the stress are $P_0$. }
\label{CompareStressHist}
\end{figure}

\begin{figure}[htp]
\centering
\includegraphics[width=1.0\hsize]{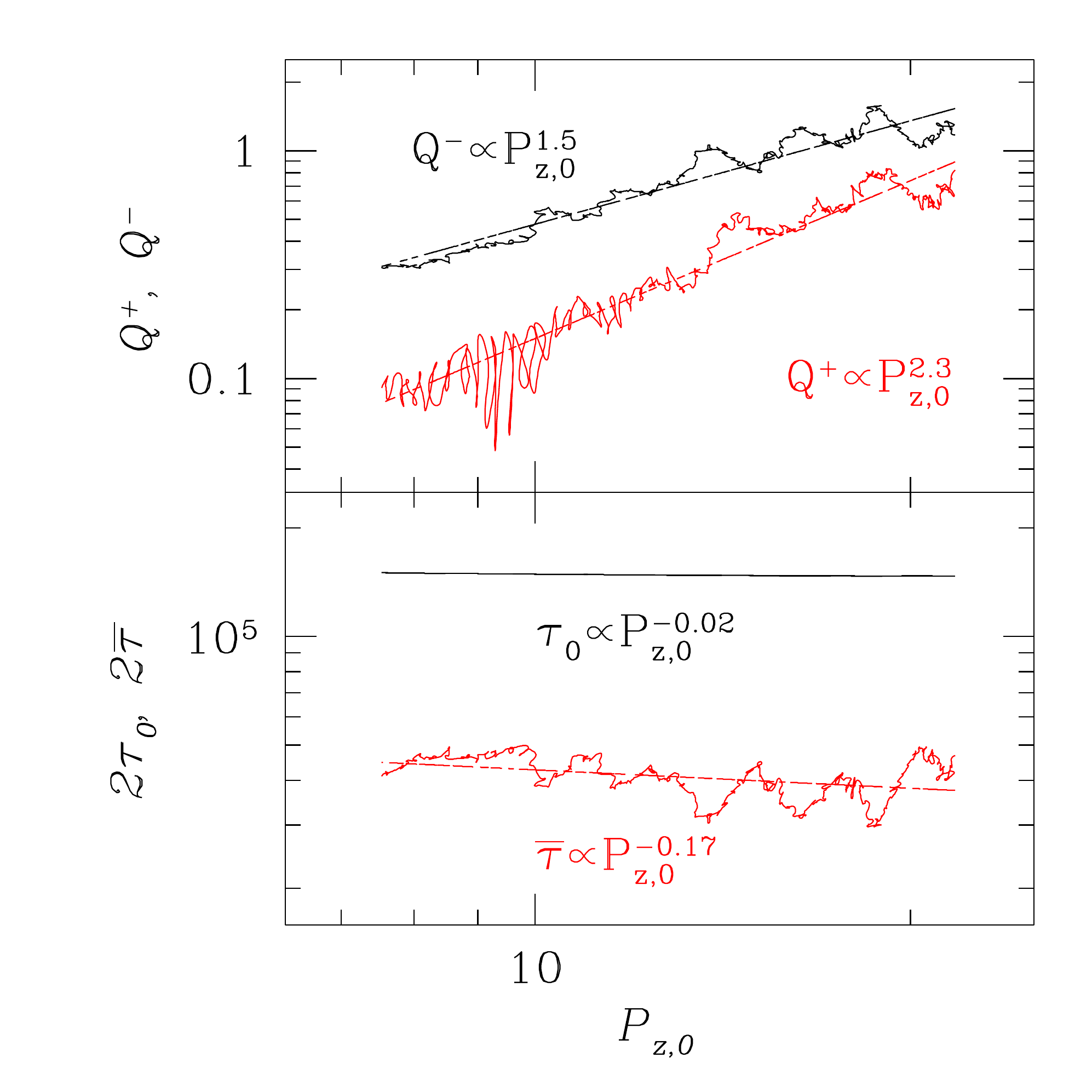}
\caption{Change of the heating $Q^+$, cooling $Q^-$ rates per unit area (top) 
and total optical depth $\tau_0$ as well as flux weighted optical depth 
$\bar{\tau}$ (bottom) as a function of the midplane total pressure $P_{z,0}$ 
for the run {\sf ESR20c}. The units are the same as in Figure \ref{HeatingCooling}. }
\label{CompareHeatingCooling}
\end{figure}

Histories of the volume averaged energy densities  and Maxwell stress of the three runs are shown in Figure \ref{CompareEnergyHist} 
and \ref{CompareStressHist}. For all the three cases, the disks continue to cool down and collapse within $\sim4-6$ thermal time scales. 
For {\sf ESR20a} and {\sf ESR20b} 
where we start the simulations from the laminar state, they collapse more quickly because there is no heating at the beginning. The Maxwell 
stress reaches the peak within the initial $\sim 6$ orbits and declines while the disks collapse. They do not reach any radiation 
pressure dominated thermal equilibrium state as the run {\sf OPALR20}. Instead, they behave in a very similar way as the simulation 
RSVET shown in \cite{Jiangetal2013c}. For the run {\sf ESR20c} where heating from the MRI turbulence exists from the beginning, 
the initial radiation energy density of the disk is pretty close to value as predicted by the radiation pressure dominated standard thin disk solution with the same 
surface density and an equivalent $\alpha=0.02$. However, with only electron scattering and free-free opacities, 
the disk does not adjust itself to reach a radiation pressure dominated equilibrium state. Instead, 
$E_r$ drops by one order of magnitude continuously within $30$ orbits. The Maxwell stress also decreases while the disk collapses. 
The dependences of heating $Q^+$ and cooling $Q^-$ rate on the midplane pressure for the run {\sf ESR20c} are shown in the top panel 
of Figure \ref{CompareHeatingCooling}, which are very similar to the simulation {\sc RMLVET} reported by \cite{Jiangetal2013c}. The heating rate 
does have a stronger sensitivity to the midplane pressure compared with the cooling rate when the thermal runaway happens. Compared with {\sf OPALR20}, the three 
runs confirm that the different behaviors are indeed caused by the iron opacity bump, as the opacity law is the only difference between them.

\subsection{Vertical Structure of the Disk in the Run {\sf OPALR20}}
\label{VerticalStructure}

In order to investigate the reasons why the iron opacity bump can make the 
radiation pressure dominated disks last much longer, we first study the time averaged 
vertical structures of disk in the run {\sf OPALR20}. We compute time averages starting
at 60 orbits so that only the VET portion of the runs is included.

\begin{figure}[htp]
\centering
\includegraphics[width=1.0\hsize]{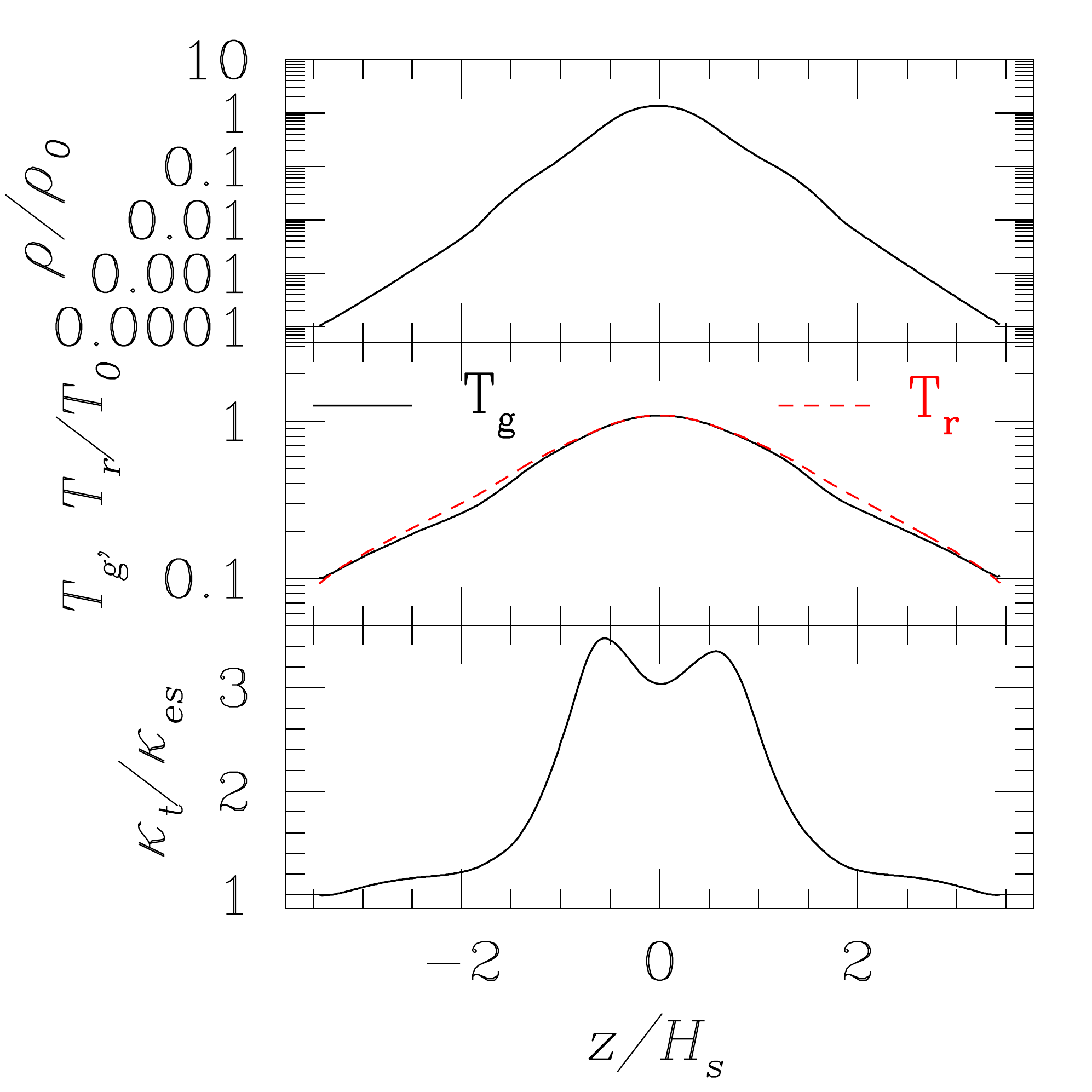}
\caption{Time and horizontally averaged vertical profiles of density $\rho$ (top panel), gas ($T_g$) and 
radiation ($T_r$) temperature (middle panel),  as well as the Rosseland mean opacity $\kappa_t$ (bottom panel) 
for the run {\sf OPALR20}.  }
\label{rhoTprofile}
\end{figure}

The horizontally averaged vertical profiles of $\rho$, $T_g$, $T_r$ and $\kappa_t$ are shown in Figure \ref{rhoTprofile}. 
The midplane temperature is larger than $T_0$ so that the peak of the Fe opacity bump occurs off the midplane.
The opacity $\kappa_t$ peaks around $\approx 0.5H_s$ and drops both when the temperature 
decreases towards the photosphere and when it increases towards the midplane. At the peak, $\kappa_t$ is more than three times the 
value of the electron scattering opacity for the solar metallicity we adopt. The iron opacity has a very weak dependence on density 
(Figure 2 of \citealt{Jiangetal2015}). The rapid drop around $\pm H_s$ is primarily because of the drop in temperature. 
 
\begin{figure}[htp]
\centering
\includegraphics[width=1.0\hsize]{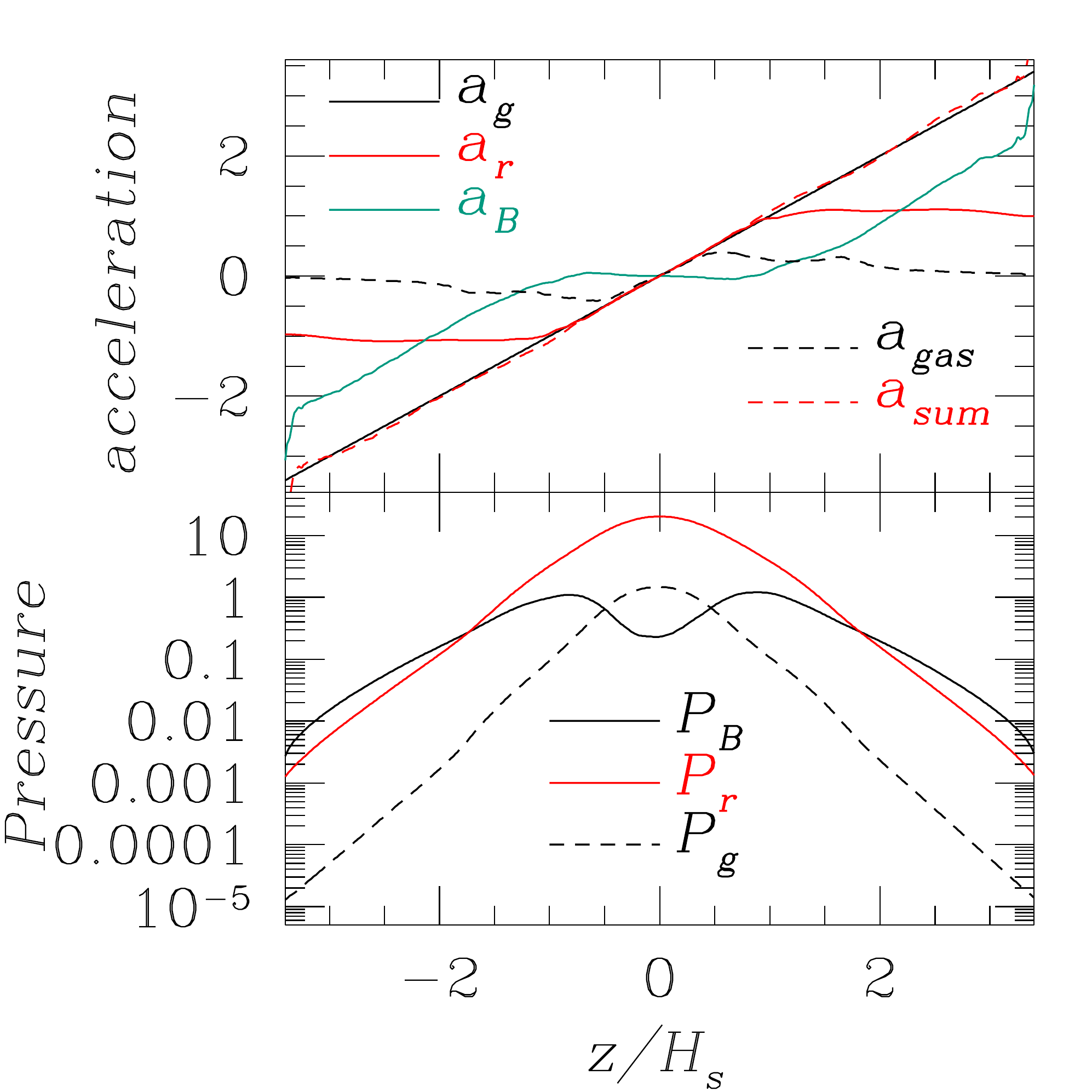}
\caption{Top: time and horizontally averaged vertical profiles of accelerations due to gas $a_{gas}$, radiation 
$a_r$ and magnetic field $a_B$ for the run {\sf OPALR20}. The sum of the three accelerations is $a_{sum}$ 
while $a_g$ is the vertical component of the 
gravitational acceleration due to the central black hole. Bottom: averaged vertical profiles of 
gas pressure $P_g$, radiation pressure $P_r$ 
and magnetic pressure $P_B$. Units for the accelerations are $\Omega^2H_s$ while the pressure units are $P_0$. }
\label{ForceBalance}
\end{figure}

Because local dynamic time scale $1/\Omega$ is much shorter than the thermal time scale, we expect the 
hydrostatic equilibrium to be maintained very well. This means the time averaged vertical accelerations due to 
various forces should be roughly balanced as 
\begin{eqnarray}
a_g=a_r+a_{gas}+a_B,
\label{eq:force}
\end{eqnarray}
where $a_g=\Omega^2z$ is the vertical component of the gravitational acceleration due to the central black hole. 
The accelerations due to radiation ($a_r$), gas ($a_{gas}$) and magnetic field ($a_B$) are calculated as
\begin{eqnarray}
a_r&=&\frac{\kappa_tF_{r,z0}}{c},\nonumber \\
a_{gas}&=&-\frac{\partial P_g}{\rho\partial z}, \nonumber\\
a_B&=&-\frac{1}{2\rho}\frac{\partial }{\partial z}\left(B_x^2+B_y^2\right)\nonumber\\
&+&\frac{1}{\rho}\left(B_x\frac{\partial B_z}{\partial x}
+B_y\frac{\partial B_z}{\partial y}\right).
\end{eqnarray}
Here $B_x$ and $B_z$ are the radial and vertical components of the magnetic field and 
$a_B$ includes both the accelerations due to magnetic pressure gradient and vertical 
component of magnetic tension \citep[][]{Blaesetal2011}. Notice that only the diffusive radiation 
flux $F_{r,z0}$ contributes to the radiation acceleration $a_r$. The time averaged vertical profiles 
of these accelerations are shown in the top panel of Figure \ref{ForceBalance}, which shows that 
equation (\ref{eq:force}) is indeed satisfied very well. 
The vertical profiles of radiation pressure $P_r$, gas pressure $P_g$ and magnetic pressure $P_b$ are shown 
in the bottom panel of Figure \ref{ForceBalance}.
Near the disk midplane when $\left|z\right|\lesssim1.8H_s$, the disk is radiation pressure 
dominated, which is also consistent with the fact $a_r$ is much larger than $a_{gas}$ and $a_B$ in this region. 
Beyond that, magnetic pressure becomes dominant. The magnetic pressure first 
increases with distance near the disk midplane and peaks 
around $\pm 0.7H_s$. Then it drops with height. This feature is commonly observed in previous 
MRI simulations \citep[e.g.,][]{MillerStone2000,Hiroseetal2006, Blaesetal2011, Jiangetal2014}. 
The ratio between radiation pressure and gas pressure varies from $\sim 15$ to more than $100$. 

\subsubsection{Change of the Optical Depth with Temperature}
Because of the sensitive dependence of the iron opacity bump with temperature, it will only 
enhance the opacity significantly within a narrow temperature range, which corresponds to a 
certain vertical range of the disk. The contribution of the iron opacity bump to the total optical depth 
is dominated by $\tau_p=\rho_p \kappa_pH_{p}$, where $\rho_p$ and $\kappa_p$ are the density and 
opacity at the opacity peak while $H_p$ is proportional to the disk scale height. For a constant surface density, 
this can be rewritten as $\tau_p=\rho_p\kappa_p\Sigma/\rho_{z,0}$, where $\rho_{z,0}$ is the density at disk midplane. 
When the disk becomes 
hotter and midplane temperature increases, the iron opacity peak will move to larger height and thus 
$\rho_p/\rho_{z,0}$ decrease. When the disk becomes colder and midplane temperature drops, the iron opacity peak will move 
towards the midplane and $\rho_p/\rho_{z,0}$ increases. Because $\kappa_p$ only depends on 
density weakly, as long as the whole iron opacity bump is included in the disk and it is the dominant opacity, this means 
the total optical depth will decrease when the disk becomes hotter, and increase with the disk becomes colder. 

To confirm this, we plot the total optical depth $2\tau_0$ as a function of the midplane pressure (dominated by $P_r$) 
for the run {\sf OPALR20} after the initial $60$ orbits at the bottom panel of Figure \ref{HeatingCooling}, which shows a very 
clear anti-correlation between $\tau_0$ and $P_{z,0}$. A simple power law fitting shows that $\tau_0\propto P_{z,0}^{-0.89}$ 
in this simulation. This is completely different from the electron scattering dominated case as shown in the bottom panel of Figure 
\ref{CompareHeatingCooling} for the run {\sf ESR20c}, where $\tau_0$ is almost independent of $P_{z,0}$ as expected. 
This is critical for the thermal instability, because in the standard thin disk model, the cooling rate $Q^-\propto P_{z,0}/\tau_0$. 
With this close to linear anti-correlation between $\tau_0$ and $P_{z,0}$, the sensitivity of the cooling rate to midplane pressure
is enhanced. This is one important reason why the thermal instability is suppressed and the disk can last much longer. 

One crude way to understand the anti-correlation between $\tau_0$ and $P_{z,0}$ quantitively is by assuming a constant ratio 
between radiation pressure and gas pressure. This is expected with efficient convection in the radiation pressure dominated 
regime so that the radiation entropy per unit mass, which is proportional to $P_r/P_g$, is roughly a constant. This is clearly 
not the case for the whole disk as this ratio increases from $\sim15$ near the disk midplane to more than $100$ near the surface. 
However, in the region when $\left|z\right|\lesssim 0.5H_s$, $P_r/P_g$ does not vary too much and this is  not a very bad assumption. Once we adopt 
this assumption, we have $\rho_p/\rho_{z,0}=\left(T_p/T_{z,0}\right)^3$, where $T_{z,0}$ is the midplane temperature. 
Because the iron opacity peak only exists for a roughly 
fixed temperature $T_p$ and $P_{z,0}\propto T_{z,0}^4$ in the radiation pressure dominated regime, the total optical depth 
contributed by the iron opacity peak can be roughly estimated as $\tau_p\propto P_{z,0}^{-0.75}$, which is actually pretty close 
to what we get from the simulation. 

\subsubsection{Vertical Energy Transport}

\begin{figure}[htp]
\centering
\includegraphics[width=1.0\hsize]{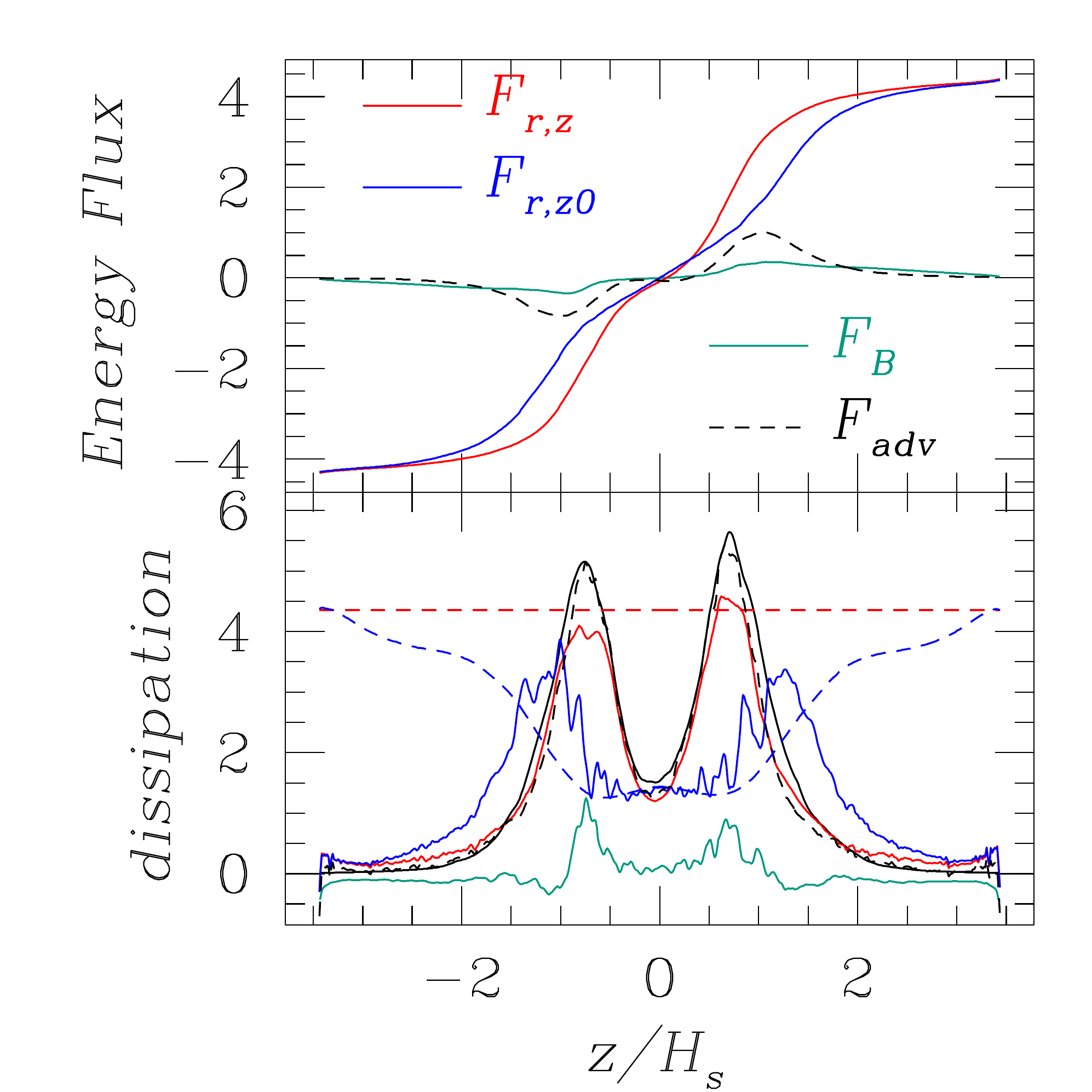}
\caption{Top: time and horizontally averaged vertical profiles of lab frame radiation flux $F_{r,z}$, 
diffusive radiation flux $F_{r,z0}$, advective radiation flux $F_{adv}$ and Poynting flux $F_B$. 
Bottom: averaged vertical profiles of local dissipation rate. The solid black line is the work done 
by the stress at each height $q\Omega(-B_xB_y+\rho v_x\delta v_y)$. The solid red, blue and green 
lines are $dF_{r,z}/dz$, $dF_{r,z0}/dz$ and $dF_B/dz$, respectively. The dashed black line is the sum of 
$dF_{r,z}/dz$ and $dF_B/dz$. The dashed red line is the critical dissipation rate $c\Omega^2/\kappa_{\rm es}$, 
while the dashed blue line is $c\Omega^2/\kappa_t$. 
Units for the energy flux and dissipation are $c_gP_0$ and $c_gP_0/H_s$. }
\label{dissipation}
\end{figure}

As shown in the top panel of Figure \ref{HeatingCooling}, during the thermal equilibrium state, the total heating 
rate $Q^+$ is balanced by the total cooling rate $Q^-$, where $Q^-$ is dominated by radiative cooling. In the 
standard thin disk model, only the diffusive radiation flux is considered for the vertical radiative cooling everywhere. 
This is not necessary the case as photons can also be advected out with the buoyantly 
rising fluid elements \citep[][]{Blaesetal2011, Jiangetal2014c}. The total energy flux carried by the photons is the lab frame 
radiation flux $F_{r,z}$, which is the sum of the diffusive radiation flux $F_{r,z0}$ and radiation enthalpy flux
\begin{eqnarray}
F_{r,z}=F_{r,z0}+v_zE_r+\left.\bv\cdot{\sf P_r}\right|_{z}.
\end{eqnarray}
Here $\bv$ is the fluid velocity and ${\sf P_r}$ is the radiation pressure tensor. The net advection flux is 
\begin{eqnarray}
F_{adv}=v_zE_r.
\end{eqnarray}
Another important component of energy flux is the Poynting flux
\begin{eqnarray}
F_B=B^2v_z-\left(\bb\cdot\bv\right)B_z.
\end{eqnarray}
The mechanical energy flux and gas enthalpy flux are small and we will not show them here. Time averaged vertical 
profiles of these energy fluxes are shown in the top panel of Figure \ref{dissipation}. The local dissipation rates 
at each height are simply $q^-_r=dF_{r,z}/dz$, $dF_{r,z0}/dz$, $dF_{adv}/dz$ and $q^-_B=dF_B/dz$. The work done by the stress 
at each height is $q^+=q\Omega\left(-B_xB_y+\rho v_x\delta v_y\right)$. These local heating and cooling rates are shown 
in the bottom panel of Figure \ref{dissipation}. In the thermal equilibrium state,  local energy conservation requires
\begin{eqnarray}
q^+=q^-_r+q^-_B.
\end{eqnarray}
Time averaged vertical profiles of the left and right hand sides are shown as the solid and dashed black lines in the bottom panel 
of Figure \ref{dissipation}, which shows that they do agree very well. 

Around $z=\pm H_s$, there is a big difference between $F_{r,z}$ and $F_{r,z0}$, because there is a significant advection flux $F_{adv}$, 
which is comparable to $F_{r,z0}$. The nature of the advection flux will be discussed in the next section. The importance of the advection 
flux was first pointed out by \cite{Hiroseetal2009} and further explored in later work (\citealt{Blaesetal2011,Jiangetal2013c}). Significant
advective transport is also observed in the global radiation MHD 
simulation of super-Eddington accretion disk by \cite{Jiangetal2014c}. The advection flux 
drops around $\pm 2H_s$ as energy is increasingly transported by the diffusive flux. Because the 
advection flux transports energy from the region near the midplane of the disk to the part closer to the surface in spite of the large optical depth, 
only the dissipation rate $dF_{r,z}/dz$ follows the heating rate $q^+$ and the vertical distribution of $dF_{r,z0}/dz$ is much broader. 
The peak of the heating rate $q^+$ is offset from the midplane 
and consistent with the peak of the magnetic pressure $P_B$ shown in the bottom panel of Figure \ref{ForceBalance}. 
The Poynting flux also peaks at the same location as $q^+$ and drops quickly with height.

\subsubsection{Nature of the Advection Flux}
\label{sec:advection}

\begin{figure*}[htp]
\centering
\includegraphics[width=1.0\hsize]{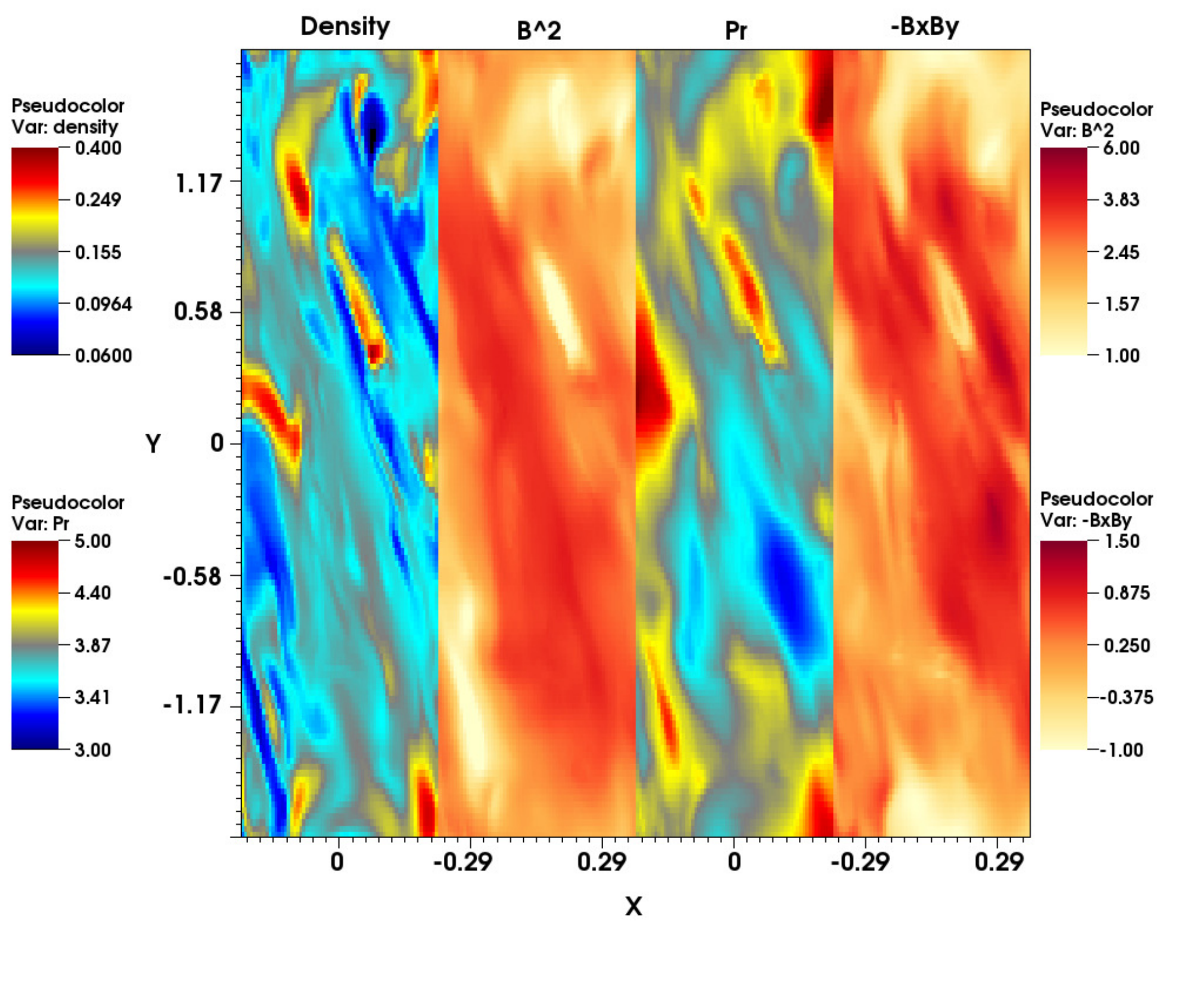}
\caption{Horizontal slice of density, magnetic pressure, radiation pressure and Maxwell stress at 
$z=H_s$ for the run {\sc OPALR20} at $100$ orbit. }
\label{Slice}
\end{figure*}

The advection flux plays an important role for the stability and structure of the disk. Particularly with the iron opacity bump, 
it is significantly enhanced with respect to the electron scattering case for the reasons we will explain now. 
The advection flux is associated with the buoyantly rising pattern of the butterfly diagram shown in Figure \ref{OPALR20STplot}. 
Around $z=\pm H_s$, the local optical depth per typical scale height $H_s$ is $\tau=\rho \kappa_tH_s=3.77\times 10^4$, which is larger than the critical optical 
depth $\tau_c\equiv c/c_g=5.69\times 10^3$ defined in \cite{Jiangetal2015}. According to the criterion of efficient radiation pressure dominated convection studied by 
\cite{Jiangetal2015}, the photon diffusion time scale is much longer than any local dynamic time scale when $\tau>\tau_c$ 
so that photons can be trapped and advected with the buoyantly rising fluid elements.  Around $\pm 2H_s$, the local optical depth $\tau$ drops to $7.61\times 10^2$, which 
is much smaller than $\tau_c$. The diffusion time becomes short and the photons cannot be trapped anymore. The photons are released and transported out 
by the diffusive flux.  The fluid elements must eventually fall back (on average) since we do not observe a net outward mass flux when average over dynamo cycles. 
However, the falling fluid 
elements are (on average) colder compared with rising elements because radiative energy is lost due to photon diffusion near the surface. Thus, there is a net vertical advective 
energy flux. Since the advection flux is much less sensitive to the optical depth, the net radiation flux  at the surface 
of the disk can becomes much larger for a given $P_{z,0}$ than the value $\sim cP_{z,0}/\tau_0$  predicted by the diffusion equation. 
%This is also the reason why the radiation efficiency can be increased in the simulation of \cite{Jiangetal2014c}.  I think this is a bit out of place here --SWD

\begin{figure}[htp]
\centering
\includegraphics[width=1.0\hsize]{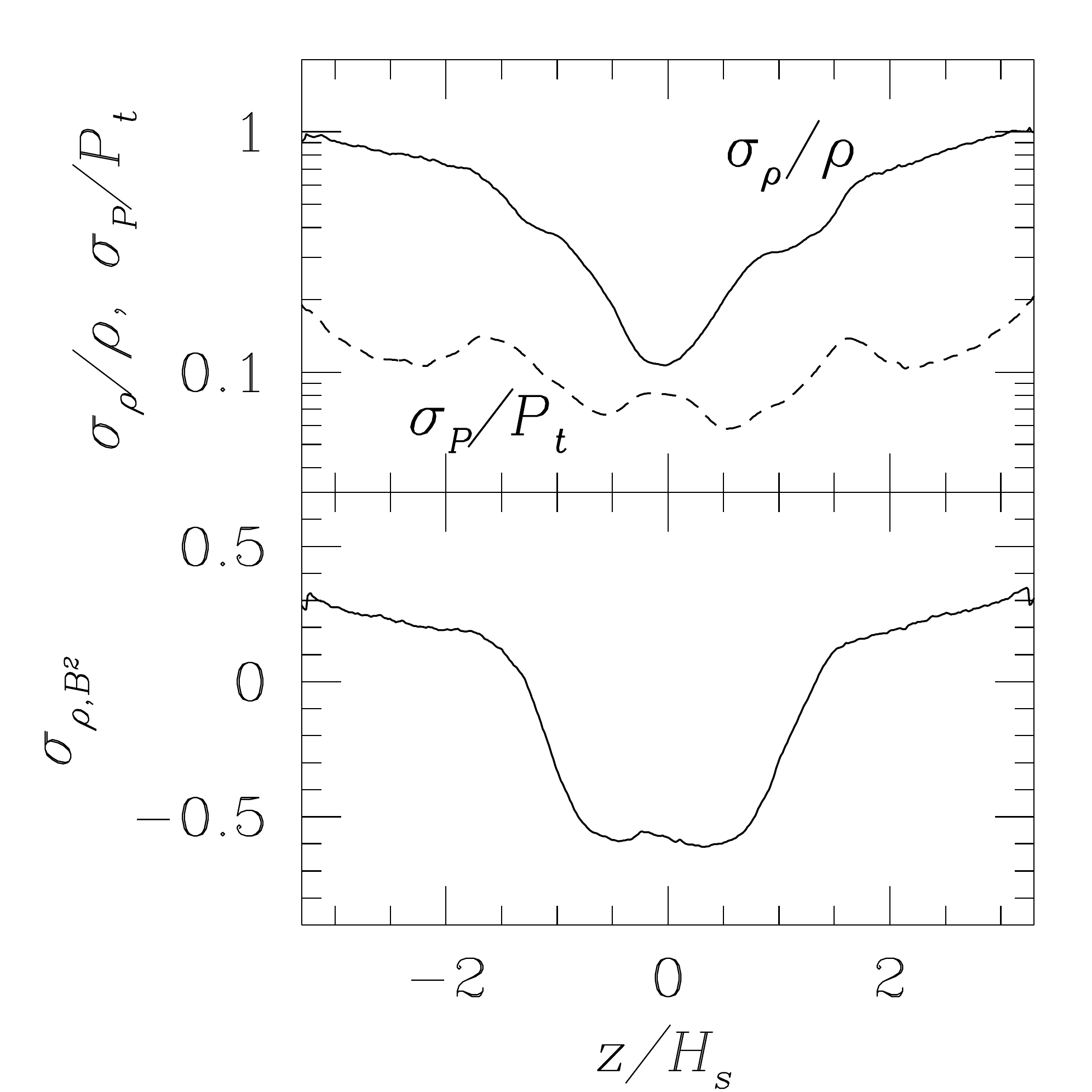}
\caption{Top: time averaged vertical profiles of the standard deviations of density $\rho$ and 
total pressure $P_t$ scaled with the horizontally averaged values for the run {\sf OPALR20}. 
Bottom: time averaged vertical profile of the cross correlation between $\rho$ and $B^2$. }
\label{Fluctuation}
\end{figure}

\cite{Blaesetal2011} provide a detailed study on why the fluid elements rise buoyantly. Buoyant motion is {\it not} driven by the standard convective instability because
the entropy profile is stable even when radiation entropy is included. 
 Instead, it is the nonlinear outcome of the anti-correlation between density and magnetic field fluctuations in the MRI turbulence. 
In the radiation pressure dominated regime, large density and magnetic field fluctuations can be produced by MRI turbulence \citep[][]{Turneretal2003,Jiangetal2013b}. 
If we consider horizontal variations at fixed height,
low density regions have larger magnetic pressure to maintain the pressure balance horizontally. 
Therefore, low density, highly magnetized regions rise buoyantly (and vice versa). We have confirmed the presence of this effect in the run {\sf OPALR20}. Figure \ref{Slice} shows the horizontal slice at $z=H_s$ of density, 
magnetic pressure, radiation pressure and Maxwell stress at $100$ orbit. The relatively lower density regions have stronger magnetic pressure, and also relatively lower 
radiation pressure to maintain the horizontal pressure balance. The Maxwell stress is well correlated with the magnetic pressure. The fluctuations can be quantified 
by the standard deviations, while the anti-correlation between density and magnetic pressure can be quantified by the cross correlation coefficient
\begin{eqnarray}
\sigma_{\rho, B^2}=\frac{\langle\left(\rho-\overline{\rho}\right)\left(B^2-\overline{B^2}\right)\rangle}{\sigma_{\rho}\sigma_{B^2}},
\end{eqnarray}
where $\langle\cdot\rangle$ means horizontal average of the quantity and $\overline{\rho}$ and $\overline{B^2}$ are the horizontally averaged $\rho$ and $B^2$ at each height. 
The standard deviations of $\rho$ and $B^2$ are $\sigma_{\rho}$ and $\sigma_{B^2}$.  The top panel of Figure \ref{Fluctuation} shows the vertical profiles of the 
standard deviations of $\rho$ and $P_t\equiv P_B+P_r+P_g$ scaled with the horizontally averaged $\rho$ and $P_t$. The time averaged vertical profile of 
$\sigma_{\rho,B^2}$ is shown at the bottom panel of Figure \ref{Fluctuation}. It is clear that density fluctuations are much larger than the fluctuations of total pressure. 
The relative fluctuation $\sigma_{\rho}/\rho$ can reach $30\%$ at $\pm H_s$. There is a strong anti-correlation between density and magnetic pressure fluctuations as 
$\sigma_{\rho, B^2}$ is negative for $\left|z\right|\lesssim 1.5H_s$.

The total flux $F_{\rm max}$ we get from the simulation only corresponds to $\sim 2\%-3\%$ Eddington accretion rate defined based on the electron scattering opacity. 
But we already observe significant advection flux in this simulation with iron opacity. For the comparison runs {\sf ESR20a}, {\sf ESR20b} and {\sf ESR20c}, advection flux 
is negligible everywhere. For the electron scattering dominated simulations presented in \cite{Jiangetal2013c}, the advection flux only became comparable to the local 
diffusive flux when the total radiation flux was 
more than $10\%$ Eddington luminosity. Therefore, it seems that we are seeing an enhancement of the advection flux in the simulation with the additional Fe opacity.

One reason for the enhanced advective flux is that the higher opacity requires it. The amount of energy that can be transported by the diffusive flux is actually 
constrained by the hydrostatic equilibrium, because in the radiation pressure dominated regime $F_{r,z0}=c\Omega^2z/\kappa_t$ and $dF_{r,z0}/dz=c\Omega^2/\kappa_t$. 
However, the local heating rate from the MRI turbulence $q^+$ is not limited by the opacity. As shown in the bottom panel of Figure \ref{dissipation}, $q^+$ is much larger 
than $dF_{r,z0}/dz$. In order to maintain a thermal equilibrium state, the excess energy must be carried out by advection. 
Notice that $dF_{r,z0}/dz$ is significantly below the red dashed curved corresponding to $c\Omega^2/\kappa_{\rm es}$ in Figure \ref{dissipation}
because $\kappa_t > \kappa_{\rm es}$.  Comparison with Figure 2 of \citet{Blaesetal2011} shows that $dF_{r,z0}/dz$ must instead follow
the $c\Omega^2/\kappa_{\rm es}$ curve in electron scattering dominated simulations, at least for the inner few scale heights.  Since $dF_{r,z0}/dz$, and therefore $F_{r,z0}$, 
must be higher in this electron scattering dominated case, there is less need for advective flux to transport energy.
%If the opacity is electron scattering $\kappa_{\rm es}$ as indicated by the dashed red line in the bottom panel of Figure \ref{dissipation}, $dF_{r,z0}/dz$ is much larger 
%because of the drop of opacity and $q^+$ is only slightly above $dF_{r,z0}/dz$ at the peak. Therefore, the thermal equilibrium can be easily maintained with hydrostatic 
%equilibrium without much advection flux. 

\begin{figure}[htp]
\centering
\includegraphics[width=1.0\hsize]{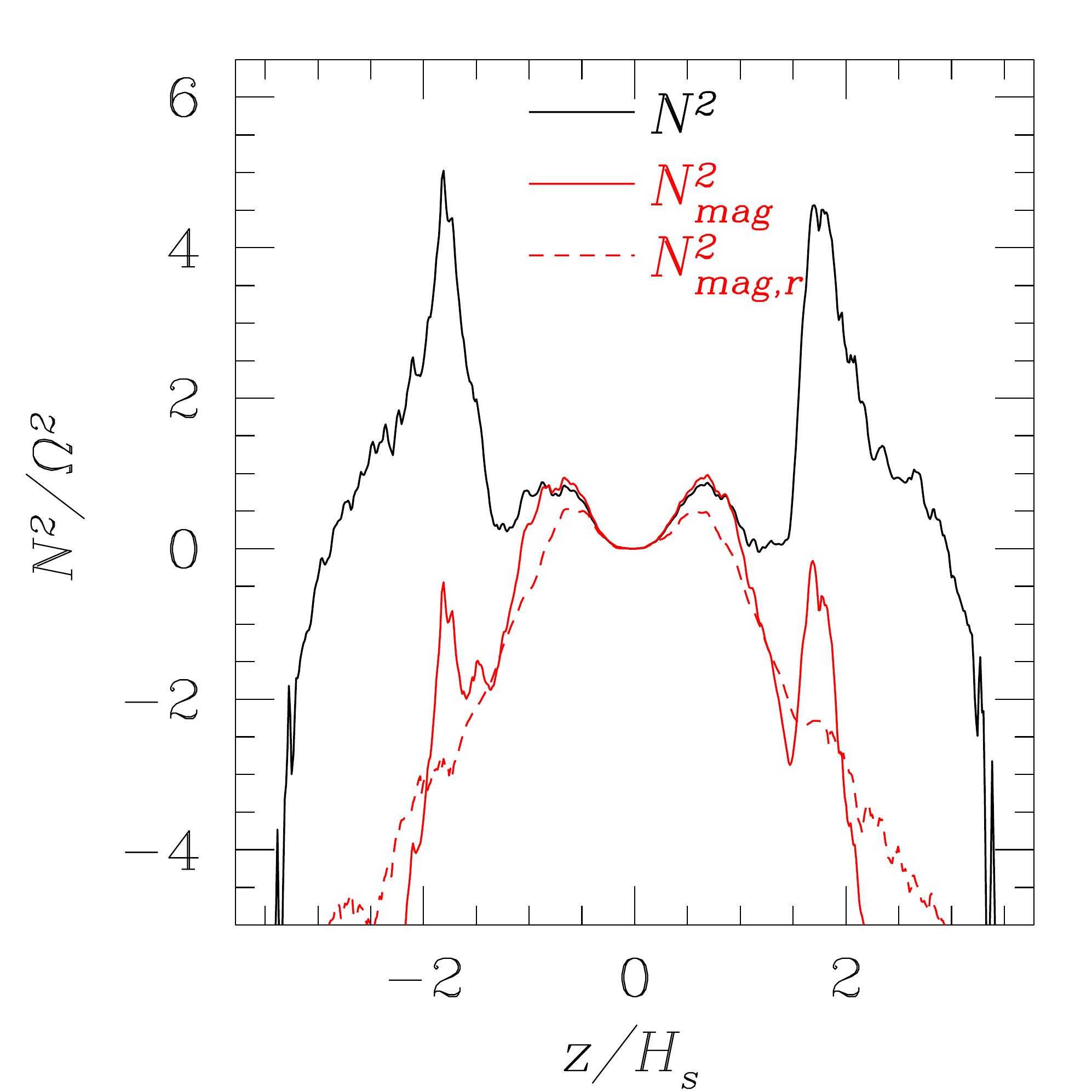}
\caption{Time averaged vertical profiles of the hydrodynamic and  
magnetohydrodynamic Brunt-V\"ais\"al\"a frequencies $N^2$, $N_{\rm mag}^2$ and $N_{\rm mag,r}^2$ 
for the run {\sf OPALR20}.}
\label{BVfrequency}
\end{figure}

It also seems to be the case that the simulations with Fe opacity are less stable to magnetic buoyancy instabilities.
We can considier this by comparing the hydrodynamic and magnetohydrodynamic Brunt-V\"ais\"al\"a frequencies as done by \cite{Blaesetal2011}. 
Following equations (27), (28) and (29) in \cite{Blaesetal2011}, the square of the hydrodynamic Brunt-V\"ais\"al\"a frequency is 
\begin{eqnarray}
N^2=a_g\left(\frac{1}{\Gamma_1}\frac{d\ln\left(P_r+P_g\right)}{dz}-\frac{d\ln \rho}{dz}\right),
\end{eqnarray}
where $\Gamma_1$ is the first adiabatic index defined in equation (15) of \cite{Jiangetal2015}. Negative $N^2$ will indicate that it is unstable to convection.  
In the regime when the diffusion time scale is longer than the local dynamic time scale, the magnetic Brunt-V\"ais\"al\"a frequency can be defined as
 \begin{eqnarray}
N_{\rm mag}^2=N^2+\frac{a_gv_A^2}{c_t^2}\frac{d\ln B}{dz},
\end{eqnarray}
where $v_A$ is the Alfv\'en velocity and $c_t^2\equiv \Gamma_1\left(P_r+P_g\right)/\rho$.
When the diffusion time scale is much shorter than the local dynamic time scale so that diffusion is rapid, the relevant magnetic Brunt-V\"ais\"al\"a frequency is
 \begin{eqnarray}
N_{\rm mag,r}^2=\frac{a_gv_A^2}{c_t^2+v_A^2}\frac{d\ln B}{dz}.
\end{eqnarray}
Undulatory Parker instability will happen when $N_{\rm mag}^2$ or $N_{\rm mag,r}^2$ become negative in the appropriate regime. Time averaged vertical profiles of 
$N^2$, $N_{\rm mag}^2$ and $N_{\rm mag,r}^2$ are shown in Figure \ref{BVfrequency}. Here we first calculate the Brunt-V\"ais\"al\"a frequencies based on the 
horizontally averaged vertical profiles at each snapshot and then time average them.
Figure \ref{BVfrequency} shows the averaged vertical profiles of $N^2$, $N_{\rm mag}^2$ and $N_{\rm mag,r}^2$. Consistent with 
\cite{Blaesetal2011}, $N^2$ is positive so that the disk is stable to hydrodynamic convection. However, $N^2$ first increases with height within $\pm 0.7H_s$ 
and then drops and reaches a minimum $\sim 0.03\Omega^2$ around $\pm 1.2-1.5H_s$. This minimum does not exist in the electron scattering case shown in \cite{Blaesetal2011}
and it is caused by the iron opacity peak around $\pm 0.7H_s$ (see the bottom panel of Figure \ref{rhoTprofile}). 

Because magnetic pressure also peaks near  $\pm0.7H_s$ and decreases with height after the peak, it makes the magnetic Brunt-V\"ais\"al\"a frequency 
$N^2_{\rm mag}$ to become negative around $\pm H_s$ as shown in Figure \ref{BVfrequency}, although magnetic pressure is still much smaller than the radiation pressure 
there. This means the region that is unstable to Parker instability is much deeper compared with the electron scattering opacity dominated case. 
Above $\pm 1.5H_s$, diffusion is rapid and the relevant magnetic Brunt-V\"ais\"al\"a frequency is $N_{\rm mag,r}^2$, which is also negative. This magnetic pressure dominated 
region is Parker unstable, as pointed out by \cite{Blaesetal2011}. Because photons and gas are tightly coupled around $\pm H_s$ ($\tau>\tau_c$), the Parker instability 
enhances the buoyancy and  theadvection flux as shown in Figure \ref{dissipation}. Above $\pm 1.5H_s$ because of rapid diffusion, even though $N^2$ and $N_{\rm mag,r}^2$ 
are negative, they do not cause any significant advection flux as in the inefficient convection case studied by \cite{Jiangetal2015}. 

\section{Discussions and Conclusions} 
\label{sec:discussion}
By including the iron opacity bump in 3D local shearing box radiation MHD simulation of AGN disks, 
we show that the radiation pressure dominated accretion disks can survive many thermal time scales 
without showing significant thermal runaways. In contrast, if we just change the opacity to be electron 
scattering plus free-free as in the standard $\alpha$ disk model, the disk collapses quickly. The iron 
opacity bump can make the radiation pressure dominated accretion disks more stable because it causes 
the total optical depth to anti-correlate with the midplane pressure, and it enhances the advective cooling. 
Both effects increases the cooling rate and make the cooling rate change as fast as the heating rate.   
These results have a wide range of implications for AGN observations. 

\subsection{Comparisons with the $\alpha$ disk model}

%We have shown that the iron opacity bump has two important effects on the thermal stability and structures of the radiation 
%pressure dominated accretion disks. It leads to an anti-correlation between the total optical depth and midplane radiation 
%pressure that is absence in electron scattering dominated disks. It also enhances the advection flux, which significantly 
%affects the vertical structures of the disk.   
Neither of two effects we have identified are included in the standard thin disk model.
Although we only have solutions for one set of parameters here, we expect the advection flux will also increase 
with increasing total pressure, because the amount of energy that can be transported by the diffusive flux is limited by the 
hydrostatic equilibrium. The excess heating will be transported by the advection flux 
to maintain thermal equilibrium. 

Steady state vertical structure of a radiation pressure dominated disk is constrained by the equations of hydrostatic equilibrium
\begin{eqnarray}
\frac{dP_r}{dz}=-\rho\Omega^2z,
\end{eqnarray}
and thermal equilibrium
\begin{eqnarray}
\alpha P_{z,0} H_{\rho}\Omega=F_{\rm max}.
\end{eqnarray}
Here we adopt the same $\alpha$ ansatz as in the standard thin disk model that the 
vertical integrated stress is related to the midplane pressure $P_{z,0}$\footnote{Note that $\alpha$ defined this way is the
ratio of the vertically integrated stress to the product of midplane pressure and the density scale, which can differ
from the ratio of the vertically integrated stress to the vertically
integrated pressure shown in Figure \ref{OPALR20History}.}  and density scale height 
\begin{eqnarray}
H_{\rho}\equiv \frac{1}{\Omega}\sqrt{\frac{P_{z,0}}{\rho_{z,0}}},
\end{eqnarray}
where $\rho_{z,0}$ is the midplane density. 
From the diffusion equation
\begin{eqnarray}
\frac{dP_r}{dz}=-\frac{\rho\kappa_tF_{rz,0}}{c}, 
\end{eqnarray}
we can define the flux weighted optical depth $\bar{\tau}$ as 
\begin{eqnarray}
\bar{\tau}\equiv\frac{1}{F_{\rm max}}\int_{0}^{L_z/2} \rho\kappa_t F_{r,z0}dz.
\end{eqnarray}
Then the cooling rate per unit area $Q^-=F_{\rm max}=cP_{rz,0}/\bar{\tau}$ and 
$P_{rz,0}\approx P_{z,0}$ in the radiation pressure dominated regime. 
Because $F_{rz,0}=0$ at the disk midplane where maximum density is located and 
$F_{\rm max}$ has contributions from both the diffusive and advective components, $\bar{\tau}$ 
is smaller than $\tau_0$. Advection flux also changes the vertical density profiles of the disk and it 
causes a stronger dependence of $\bar{\tau}$ on the midplane pressure compared with $\tau_0$ 
as shown at the bottom panel of Figure \ref{HeatingCooling} for {\sf OPALR20}. To separate the 
effects of the opacity change with temperature and the advection flux, we calculate
\begin{eqnarray}
\bar{\tau}_{\rm es}\equiv\frac{1}{F_{\rm max}}\int_0^{L_z/2}\rho\kappa_{\rm es}F_{r,z0}dz. 
\end{eqnarray}
For the run {\sf OPALR20}, $\bar{\tau}\propto P_{z,0}^{-2.90}, \tau_0\propto P_{z,0}^{-0.89}$ while 
$\bar{\tau}_{\rm es}\propto P_{z,0}^{-2.10}$, which confirms that the change of $\bar{\tau}$ is the combined 
effect of $\tau_0$ and $\bar{\tau}_{\rm es}$. For comparison, the bottom panel of Figure \ref{CompareHeatingCooling} 
shows that in the run {\sf ESR20c} with electron scattering and free-free opacities, 
 $\bar{\tau}$ only changes with midplane pressure as $P_{z,0}^{-0.17}$ while $\tau_0$ is almost 
a constant as expected. Notice that the midplane gas pressure is already half of the radiation pressure at the end of this run 
and $Q\neq P_{z,0}/(c\bar{\tau})$.

Compared with the classical argument of thermal instability \citep[][]{ShakuraSunyaev1976,Piran1978,Pringle1981}, 
 $Q^-\propto P_{z,0}/\tau_0$. When electron scattering opacity dominates $\tau_0$ is expected to be nearly constant 
because the surface density changes on timescale much longer than the thermal time scale. 
However, with iron opacity, we find that $Q^-\propto P_{z,0}/\bar{\tau}$ can increase much more sensitively
with $P_{z,0}$ than the standard model predicts. Since the sensitivity of $Q^+$ to $P_{z,0}$ is not significantly affected
by the additional iron opacity, $Q^+$ and $Q^-$ can have rather similar dependencies on midplane pressure.  This is consistent with
thermal runaway being significantly slower and possibly absent in {\sf OPALR20}, but occurring rapidly in the electron scattering
dominated runs.

Since we cannot determine how the effects of iron opacity peak change with different accretion rates with the available simulations, we 
parametrize $\bar{\tau}$ as 
\begin{eqnarray}
\bar{\tau}\equiv \beta \frac{\kappa_{\rm es}\Sigma}{2}.
\end{eqnarray} 
Therefore, the $\Sigma$, $P_{z,0}$ and $H_{\rho}$ are related to $F_{\rm max}$ and $\Omega$ as
\begin{eqnarray}
H_{\rho}&=&\frac{\beta\kappa_{\rm es}}{c}\frac{F_{\rm max}}{\Omega^2},\nonumber\\
P_{z,0}&=&\frac{c}{\beta\kappa_{\rm es}\alpha}\Omega,\nonumber\\
\Sigma&=&2\rho_{z,0}H_{\rho}=\frac{2c^2}{\beta^2\kappa_{\rm es}^2\alpha}\frac{\Omega}{F_{\rm max}}.
\end{eqnarray}
To connect to the radial structures of the accretion disk, the radiation flux can be related 
to the accretion rate $\dot{M}$ as \citep[][]{ShakuraSunyaev1973,Franketal2002}
\begin{eqnarray}
F_{\rm max}=\frac{3G\mbh\dot{M}}{8\pi r_0^3}.
\end{eqnarray}
If we scale the accretion rate with the Eddington accretion rate $\dot{M}_{\rm Edd}=40\pi G\mbh/\left(c\kappa_{\rm es}\right)$ as 
$\dot{m}\equiv\dot{M}/\dot{M}_{\rm Edd}$, and scale the radius with the Schwarzschild radius $r_s=2G\mbh/c^2$ as 
$r\equiv r_0/r_s$, the equations can be simplified as
\begin{eqnarray}
H_{\rho}&=&\frac{15}{2}\beta r_s\dot{m},\nonumber\\
P_{z,0}&=&\frac{\sqrt{2}}{2}\frac{c^2}{\beta\kappa_{\rm es}\alpha r_s}\frac{1}{r^{3/2}},\nonumber\\
\Sigma&=&\frac{4\sqrt{2}}{15}\frac{r^{3/2}}{\beta^2\alpha\kappa_{\rm es}\dot{m}}.
\label{scurveequation}
\end{eqnarray}
If we set $\beta=1$, we recover the scaling relations in the radiation pressure dominated $\alpha$ disk model. 
This is also similar to the analysis done in the Appendix A of \cite[][]{Hiroseetal2009b}.
The time averaged $\beta=0.4$ for the run {\sf OPALR20}. Calibrating $\beta$ for different $\dot{m}$ and $r$ 
will be the focus of our future work. 

In the $\alpha$ disk model, there is a maximum surface density that can support a thermal equilibrium state in the so-called 
S-curve \citep[][]{LightmanEardley1974,Hiroseetal2009b} and the surface density in
{\sf OPALR20} is {\it larger} than this maximum if $\beta=1$. If we simply increase the opacity in the $\alpha$ disk model to 
account for the additional iron opacity, this maximum surface density would {\it decrease}.  However, the effect of 
the enhanced advection on the flux weighted opacity $\bar{\tau}$ keeps $\beta < 1$, so that the maximum  
surface density can be consistent with the larger surface 
density in {\sf OPALR20}.

\subsection{Implications for AGN Observations}
The bolometric luminosity of most AGNs are observed to be smaller than the Eddington luminosity defined 
using only the electron scattering opacity \citep[][]{Heckmanetal2004,Kellyetal2010}. Therefore, the accretion disks of most 
AGNs are thought to be adequately described by the standard thin disk model. No significant outflow driven by the continuum radiation is expected. 
However, when the opacity is enhanced significantly by bound-bound transition of Fe, the effective radiation acceleration can be comparable or even larger 
than the gravitational acceleration. In other words, the Eddington ratio defined by the electron scattering opacity is no longer relevant. 
Therefore, similar to the case of massive stars \citep[][]{Smith2014}, radiation acceleration via the iron opacity bump has
the potential to drive significant outflow, even though the normal Eddington ratio is smaller than $1$. Since the location of the 
iron opacity bump is relatively close to the central supermassive black hole (20 $r_s$ for the simulation {\sf OPALR20}) and it will move inward 
with increasing black hole mass, the velocity of any possible outflow driven from this region could be on the order of the Keplerian 
rotation velocity, which might be much faster than the outflow driven by the line opacity in previous models \citep{Murrayetal1995,Progaetal2000}. 
This makes it a possible candidate for the launching of 
ultra-fast outflows \citep[][]{Tombesietal2010,Tombesietal2015}. The spectral energy distributions of most AGNs usually show a 
turnover around $1000$ {\r A} as discussed in the Introduction, and \cite{LaorDavis2014} suggests that significant outflow driven by UV line 
opacity might explain this. The iron opacity bump may play a dominant role in the launching of such an outflow.

The enhancement of the opacity due to iron also depends on the metallicity. The iron opacity bump should increase with increasing 
metallicity \citep[][]{Paxtonetal2011}. The simulation {\sf OPALR20} assumes solar metallicity as an example, but the metallicity in AGNs 
can be supersolar \citep[][]{HamannFerland1993,Aravetal2007,Fieldsetal2007}.  Therefore, this simulation may underestimate the role 
that the iron opacity bump plays in stabilizing the disk, modifying the vertical structure, and potentially 
driving an outflow. On the other hand, if the metallicity in some AGN disks is much smaller, the iron opacity 
bump will be weaker and thermal instability may still exist, although the growth time scale of the thermal instability may still be 
longer than the thermal time scale as predicted by the $\alpha$ disk model. This perhaps is one explanation for the recently discovered 
``change look'' AGNs \citep[][]{LaMassaetal2015,Ruanetal2015,Runnoeetal2016} but the significant change of luminosity 
within a few years does not happen for other AGNs.

Due to the sensitive dependence of the iron opacity bump on the temperature, it will only play an important role 
in a certain radial range of the accretion disks in AGNs, since the midplane temperature of the disk decreases with increasing 
radius for a fixed accretion rate. The iron opacity bump only exists around $1.8\times 10^5$ K.  Therefore, it will be absent in hotter regions
of the disk where the effective temperature is larger than this value, but may still be relevant near the midplane in regions of 
the disk where the effective temperature is much lower (e.g. $\sim 2\times10^4$ K in {\sf OPALR20}).    This 
suggests that the region where  the iron opacity bump is most important is between the innermost region where extreme UV or soft X-rays 
would be emitted and regions further out where longer wavelength UV photons are emitted.  
Since the effective Eddington ratio in the region with the iron opacity bump is larger, the height 
of the photosphere in this region will likely larger than in hotter, neighboring regions closer to the black hole where electron scattering
opacity would dominate. 
This could lead to a sharp drop in the scale height when radius decreases and provide a ``bump'' to shield the outer UV emitting disk from X-ray 
photons emitted in the inner region.  Since the outer UV photons are thought to accelerate broad absorption line outflows through line driving 
\citep[][]{Progaetal2000}, this geometry might help shield the outflows from over ionization by the X-rays \citep[][]{Higginbottometal2014}. Such
a geometry may also explain some properties of weak line quasars \citep{Luoetal2015}.
A sharp drop in the scale height as radius decreases would also provide a surface for intercepting X-ray photons emitted in the inner disk 
without the need for a large scale height for the X-ray emitting region.  This may naturally explain the level of X-ray irradiation inferred from correlated 
variability of UV and X-rays bands \citep[][]{Edelsonetal1996,Edelsonetal2015} and may even alter the radial distribution of the 
UV emission if the irradiating flux is large enough.
Ultimately, global radiation MHD simulations will be needed to understand the effects of the iron opacity bump
on the global structure of the disk to determine if these speculations are correct.
%We anticipate that we will soon be able to perform such simulations using the numerical algorithm in \cite{Jiangetal2014c} 
%in the newest version of {\sc Athena} code: {\sc Athena++}.

\section*{Acknowledgements}
Y.F.J. thanks Omer Blaes, Jeremy Goodman, Daniel Proga and Eliot Quataert 
for helpful
discussions. This work benefited from our participation in an
International Space Science Institute meeting.  This work was
supported by the computational resources provided by the NASA High-End
Computing (HEC) Program through the NASA Advanced Supercomputing (NAS)
Division at Ames Research Center; the Extreme Science and Engineering
Discovery Environment (XSEDE), which is supported by National Science
Foundation grant number ACI-1053575; and the computer cluster Rivanna
at the University of Virginia. Y.F.J. is supported by NASA through
Einstein Postdoctoral Fellowship grant number PF3-140109 awarded by
the Chandra X-ray Center, which is operated by the Smithsonian
Astrophysical Observatory for NASA under contract NAS8-03060.
S.W.D. is supported by a Sloan Foundation Research Fellowship. 
We acknowledge support from NSF grant AST-1333091  
``Collaborative Research: Black Hole Accretion Theory and
Computation Network".

\bibliographystyle{apj}
\bibliography{AGNDisk}

\begin{thebibliography}{62}
\expandafter\ifx\csname natexlab\endcsname\relax\def\natexlab#1{#1}\fi

\bibitem[{{Arav} {et~al.}(2007){Arav}, {Gabel}, {Korista}, {Kaastra}, {Kriss},
  {Behar}, {Costantini}, {Gaskell}, {Laor}, {Kodituwakku}, {Proga}, {Sako},
  {Scott}, \& {Steenbrugge}}]{Aravetal2007}
{Arav}, N., {Gabel}, J.~R., {Korista}, K.~T., {et~al.} 2007, \apj, 658, 829

\bibitem[{{Balbus} \& {Hawley}(1991)}]{BalbusHawley1991}
{Balbus}, S.~A., \& {Hawley}, J.~F. 1991, \apj, 376, 214

\bibitem[{{Blackburne} {et~al.}(2011){Blackburne}, {Pooley}, {Rappaport}, \&
  {Schechter}}]{Blackburneetal2011}
{Blackburne}, J.~A., {Pooley}, D., {Rappaport}, S., \& {Schechter}, P.~L. 2011,
  \apj, 729, 34

\bibitem[{{Blaes} {et~al.}(2011){Blaes}, {Krolik}, {Hirose}, \&
  {Shabaltas}}]{Blaesetal2011}
{Blaes}, O., {Krolik}, J.~H., {Hirose}, S., \& {Shabaltas}, N. 2011, \apj, 733,
  110

\bibitem[{{Bonning} {et~al.}(2007){Bonning}, {Cheng}, {Shields}, {Salviander},
  \& {Gebhardt}}]{Bonningetal2007}
{Bonning}, E.~W., {Cheng}, L., {Shields}, G.~A., {Salviander}, S., \&
  {Gebhardt}, K. 2007, \apj, 659, 211

\bibitem[{{Crummy} {et~al.}(2006){Crummy}, {Fabian}, {Gallo}, \&
  {Ross}}]{Crummyetal2006}
{Crummy}, J., {Fabian}, A.~C., {Gallo}, L., \& {Ross}, R.~R. 2006, \mnras, 365,
  1067

\bibitem[{{Davis} {et~al.}(2012){Davis}, {Stone}, \& {Jiang}}]{Davisetal2012}
{Davis}, S.~W., {Stone}, J.~M., \& {Jiang}, Y.-F. 2012, \apjs, 199, 9

\bibitem[{{Davis} {et~al.}(2007){Davis}, {Woo}, \& {Blaes}}]{Davisetal2007}
{Davis}, S.~W., {Woo}, J.-H., \& {Blaes}, O.~M. 2007, \apj, 668, 682

\bibitem[{{Dhanda} {et~al.}(2007){Dhanda}, {Baldwin}, {Bentz}, \&
  {Osmer}}]{Dhandaetal2007}
{Dhanda}, N., {Baldwin}, J.~A., {Bentz}, M.~C., \& {Osmer}, P.~S. 2007, \apj,
  658, 804

\bibitem[{{Done}(2014)}]{Done2014}
{Done}, C. 2014, in Suzaku-MAXI 2014: Expanding the Frontiers of the X-ray
  Universe, ed. M.~{Ishida}, R.~{Petre}, \& K.~{Mitsuda}, 300

\bibitem[{{Done} {et~al.}(2012){Done}, {Davis}, {Jin}, {Blaes}, \&
  {Ward}}]{Doneetal2012}
{Done}, C., {Davis}, S.~W., {Jin}, C., {Blaes}, O., \& {Ward}, M. 2012, \mnras,
  420, 1848

\bibitem[{{Edelson} {et~al.}(2015){Edelson}, {Gelbord}, {Horne}, {McHardy},
  {Peterson}, {Ar{\'e}valo}, {Breeveld}, {De Rosa}, {Evans}, {Goad}, {Kriss},
  {Brandt}, {Gehrels}, {Grupe}, {Kennea}, {Kochanek}, {Nousek}, {Papadakis},
  {Siegel}, {Starkey}, {Uttley}, {Vaughan}, {Young}, {Barth}, {Bentz},
  {Brewer}, {Crenshaw}, {Dalla Bont{\`a}}, {De Lorenzo-C{\'a}ceres}, {Denney},
  {Dietrich}, {Ely}, {Fausnaugh}, {Grier}, {Hall}, {Kaastra}, {Kelly},
  {Korista}, {Lira}, {Mathur}, {Netzer}, {Pancoast}, {Pei}, {Pogge},
  {Schimoia}, {Treu}, {Vestergaard}, {Villforth}, {Yan}, \&
  {Zu}}]{Edelsonetal2015}
{Edelson}, R., {Gelbord}, J.~M., {Horne}, K., {et~al.} 2015, \apj, 806, 129

\bibitem[{{Edelson} {et~al.}(1996){Edelson}, {Alexander}, {Crenshaw}, {Kaspi},
  {Malkan}, {Peterson}, {Warwick}, {Clavel}, {Filippenko}, {Horne}, {Korista},
  {Kriss}, {Krolik}, {Maoz}, {Nandra}, {O'Brien}, {Penton}, {Yaqoob},
  {Albrecht}, {Alloin}, {Ayres}, {Balonek}, {Barr}, {Barth}, {Bertram},
  {Bromage}, {Carini}, {Carone}, {Cheng}, {Chuvaev}, {Dietrich},
  {Dultzin-Hacyan}, {Gaskell}, {Glass}, {Goad}, {Hemar}, {Ho}, {Huchra},
  {Hutchings}, {Johnson}, {Kazanas}, {Kollatschny}, {Koratkar}, {Kovo}, {Laor},
  {MacAlpine}, {Magdziarz}, {Martin}, {Matheson}, {McCollum}, {Miller},
  {Morris}, {Oknyanskij}, {Penfold}, {Perez}, {Perola}, {Pike}, {Pogge},
  {Ptak}, {Qian}, {Recondo-Gonzalez}, {Reichert}, {Rodriguez-Espinoza},
  {Rodriguez-Pascual}, {Rokaki}, {Roland}, {Sadun}, {Salamanca}, {Santos-Lleo},
  {Shields}, {Shull}, {Smith}, {Smith}, {Snijders}, {Stirpe}, {Stoner}, {Sun},
  {Ulrich}, {van Groningen}, {Wagner}, {Wagner}, {Wanders}, {Welsh}, {Weymann},
  {Wilkes}, {Wu}, {Wurster}, {Xue}, {Zdziarski}, {Zheng}, \&
  {Zou}}]{Edelsonetal1996}
{Edelson}, R.~A., {Alexander}, T., {Crenshaw}, D.~M., {et~al.} 1996, \apj, 470,
  364

\bibitem[{{Elvis} {et~al.}(1994){Elvis}, {Wilkes}, {McDowell}, {Green},
  {Bechtold}, {Willner}, {Oey}, {Polomski}, \& {Cutri}}]{Elvisetal1994}
{Elvis}, M., {Wilkes}, B.~J., {McDowell}, J.~C., {et~al.} 1994, \apjs, 95, 1

\bibitem[{{Fields} {et~al.}(2007){Fields}, {Mathur}, {Krongold}, {Williams}, \&
  {Nicastro}}]{Fieldsetal2007}
{Fields}, D.~L., {Mathur}, S., {Krongold}, Y., {Williams}, R., \& {Nicastro},
  F. 2007, \apj, 666, 828

\bibitem[{{Frank} {et~al.}(2002){Frank}, {King}, \& {Raine}}]{Franketal2002}
{Frank}, J., {King}, A., \& {Raine}, D.~J. 2002, {Accretion Power in
  Astrophysics: Third Edition} (Cambridge University Press, Cambridge, UK)

\bibitem[{{Guan} {et~al.}(2009){Guan}, {Gammie}, {Simon}, \&
  {Johnson}}]{Guanetal2009}
{Guan}, X., {Gammie}, C.~F., {Simon}, J.~B., \& {Johnson}, B.~M. 2009, \apj,
  694, 1010

\bibitem[{{Hamann} \& {Ferland}(1993)}]{HamannFerland1993}
{Hamann}, F., \& {Ferland}, G. 1993, \apj, 418, 11

\bibitem[{{Hawley} {et~al.}(2011){Hawley}, {Guan}, \&
  {Krolik}}]{Hawleyetal2011}
{Hawley}, J.~F., {Guan}, X., \& {Krolik}, J.~H. 2011, \apj, 738, 84

\bibitem[{{Heckman} {et~al.}(2004){Heckman}, {Kauffmann}, {Brinchmann},
  {Charlot}, {Tremonti}, \& {White}}]{Heckmanetal2004}
{Heckman}, T.~M., {Kauffmann}, G., {Brinchmann}, J., {et~al.} 2004, \apj, 613,
  109

\bibitem[{{Higginbottom} {et~al.}(2014){Higginbottom}, {Proga}, {Knigge},
  {Long}, {Matthews}, \& {Sim}}]{Higginbottometal2014}
{Higginbottom}, N., {Proga}, D., {Knigge}, C., {et~al.} 2014, \apj, 789, 19

\bibitem[{{Hirose} {et~al.}(2009{\natexlab{a}}){Hirose}, {Blaes}, \&
  {Krolik}}]{Hiroseetal2009b}
{Hirose}, S., {Blaes}, O., \& {Krolik}, J.~H. 2009{\natexlab{a}}, \apj, 704,
  781

\bibitem[{{Hirose} {et~al.}(2014){Hirose}, {Blaes}, {Krolik}, {Coleman}, \&
  {Sano}}]{Hiroseetal2014}
{Hirose}, S., {Blaes}, O., {Krolik}, J.~H., {Coleman}, M.~S.~B., \& {Sano}, T.
  2014, \apj, 787, 1

\bibitem[{{Hirose} {et~al.}(2009{\natexlab{b}}){Hirose}, {Krolik}, \&
  {Blaes}}]{Hiroseetal2009}
{Hirose}, S., {Krolik}, J.~H., \& {Blaes}, O. 2009{\natexlab{b}}, \apj, 691, 16

\bibitem[{{Hirose} {et~al.}(2006){Hirose}, {Krolik}, \&
  {Stone}}]{Hiroseetal2006}
{Hirose}, S., {Krolik}, J.~H., \& {Stone}, J.~M. 2006, \apj, 640, 901

\bibitem[{{Iglesias} \& {Rogers}(1996)}]{iglesias96}
{Iglesias}, C.~A., \& {Rogers}, F.~J. 1996, \apj, 464, 943

\bibitem[{{Jiang} {et~al.}(2015){Jiang}, {Cantiello}, {Bildsten}, {Quataert},
  \& {Blaes}}]{Jiangetal2015}
{Jiang}, Y.-F., {Cantiello}, M., {Bildsten}, L., {Quataert}, E., \& {Blaes}, O.
  2015, \apj, 813, 74

\bibitem[{{Jiang} {et~al.}(2012){Jiang}, {Stone}, \& {Davis}}]{Jiangetal2012}
{Jiang}, Y.-F., {Stone}, J.~M., \& {Davis}, S.~W. 2012, \apjs, 199, 14

\bibitem[{{Jiang} {et~al.}(2013{\natexlab{a}}){Jiang}, {Stone}, \&
  {Davis}}]{Jiangetal2013c}
---. 2013{\natexlab{a}}, \apj, 778, 65

\bibitem[{{Jiang} {et~al.}(2013{\natexlab{b}}){Jiang}, {Stone}, \&
  {Davis}}]{Jiangetal2013b}
---. 2013{\natexlab{b}}, \apj, 767, 148

\bibitem[{{Jiang} {et~al.}(2014{\natexlab{a}}){Jiang}, {Stone}, \&
  {Davis}}]{Jiangetal2014c}
---. 2014{\natexlab{a}}, \apj, 796, 106

\bibitem[{{Jiang} {et~al.}(2014{\natexlab{b}}){Jiang}, {Stone}, \&
  {Davis}}]{Jiangetal2014}
---. 2014{\natexlab{b}}, \apj, 784, 169

\bibitem[{{Kelly} {et~al.}(2010){Kelly}, {Vestergaard}, {Fan}, {Hopkins},
  {Hernquist}, \& {Siemiginowska}}]{Kellyetal2010}
{Kelly}, B.~C., {Vestergaard}, M., {Fan}, X., {et~al.} 2010, \apj, 719, 1315

\bibitem[{{Koratkar} \& {Blaes}(1999)}]{KoratkarBlaes1999}
{Koratkar}, A., \& {Blaes}, O. 1999, \pasp, 111, 1

\bibitem[{{LaMassa} {et~al.}(2015){LaMassa}, {Cales}, {Moran}, {Myers},
  {Richards}, {Eracleous}, {Heckman}, {Gallo}, \& {Urry}}]{LaMassaetal2015}
{LaMassa}, S.~M., {Cales}, S., {Moran}, E.~C., {et~al.} 2015, \apj, 800, 144

\bibitem[{{Laor} \& {Davis}(2014)}]{LaorDavis2014}
{Laor}, A., \& {Davis}, S.~W. 2014, \mnras, 438, 3024

\bibitem[{{Lasota}(2001)}]{Lasota2001}
{Lasota}, J.-P. 2001, New Astronomy Reviews, 45, 449

\bibitem[{{Lightman} \& {Eardley}(1974)}]{LightmanEardley1974}
{Lightman}, A.~P., \& {Eardley}, D.~M. 1974, \apjl, 187, L1

\bibitem[{{Luo} {et~al.}(2015){Luo}, {Brandt}, {Hall}, {Wu}, {Anderson},
  {Garmire}, {Gibson}, {Plotkin}, {Richards}, {Schneider}, {Shemmer}, \&
  {Shen}}]{Luoetal2015}
{Luo}, B., {Brandt}, W.~N., {Hall}, P.~B., {et~al.} 2015, \apj, 805, 122

\bibitem[{{Malkan}(1983)}]{Malkan1983}
{Malkan}, M.~A. 1983, \apj, 268, 582

\bibitem[{{Merloni}(2003)}]{Merloni2003}
{Merloni}, A. 2003, \mnras, 341, 1051

\bibitem[{{Miller} \& {Stone}(2000)}]{MillerStone2000}
{Miller}, K.~A., \& {Stone}, J.~M. 2000, \apj, 534, 398

\bibitem[{{Morgan} {et~al.}(2010){Morgan}, {Kochanek}, {Morgan}, \&
  {Falco}}]{Morganetal2010}
{Morgan}, C.~W., {Kochanek}, C.~S., {Morgan}, N.~D., \& {Falco}, E.~E. 2010,
  \apj, 712, 1129

\bibitem[{{Murray} {et~al.}(1995){Murray}, {Chiang}, {Grossman}, \&
  {Voit}}]{Murrayetal1995}
{Murray}, N., {Chiang}, J., {Grossman}, S.~A., \& {Voit}, G.~M. 1995, \apj,
  451, 498

\bibitem[{{Paxton} {et~al.}(2011){Paxton}, {Bildsten}, {Dotter}, {Herwig},
  {Lesaffre}, \& {Timmes}}]{Paxtonetal2011}
{Paxton}, B., {Bildsten}, L., {Dotter}, A., {et~al.} 2011, \apjs, 192, 3

\bibitem[{{Piran}(1978)}]{Piran1978}
{Piran}, T. 1978, \apj, 221, 652

\bibitem[{{Pringle}(1981)}]{Pringle1981}
{Pringle}, J.~E. 1981, \araa, 19, 137

\bibitem[{{Proga} {et~al.}(2000){Proga}, {Stone}, \& {Kallman}}]{Progaetal2000}
{Proga}, D., {Stone}, J.~M., \& {Kallman}, T.~R. 2000, \apj, 543, 686

\bibitem[{{Ruan} {et~al.}(2015){Ruan}, {Anderson}, {Cales}, {Eracleous},
  {Green}, {Morganson}, {Runnoe}, {Shen}, {Wilkinson}, {Blanton}, {Dwelly},
  {Georgakakis}, {Greene}, {LaMassa}, {Merloni}, \& {Schneider}}]{Ruanetal2015}
{Ruan}, J.~J., {Anderson}, S.~F., {Cales}, S.~L., {et~al.} 2015,
  arXiv:1509.03634

\bibitem[{{Runnoe} {et~al.}(2016){Runnoe}, {Cales}, {Ruan}, {Eracleous},
  {Anderson}, {Shen}, {Green}, {Morganson}, {LaMassa}, {Greene}, {Dwelly},
  {Schneider}, {Merloni}, {Georgakakis}, \& {Roman-Lopes}}]{Runnoeetal2016}
{Runnoe}, J.~C., {Cales}, S., {Ruan}, J.~J., {et~al.} 2016, \mnras, 455, 1691

\bibitem[{{Shakura} \& {Sunyaev}(1973)}]{ShakuraSunyaev1973}
{Shakura}, N.~I., \& {Sunyaev}, R.~A. 1973, \aap, 24, 337

\bibitem[{{Shakura} \& {Sunyaev}(1976)}]{ShakuraSunyaev1976}
---. 1976, \mnras, 175, 613

\bibitem[{{Shull} {et~al.}(2012){Shull}, {Stevans}, \&
  {Danforth}}]{Shulletal2012}
{Shull}, J.~M., {Stevans}, M., \& {Danforth}, C.~W. 2012, \apj, 752, 162

\bibitem[{{Smith}(2014)}]{Smith2014}
{Smith}, N. 2014, \araa, 52, 487

\bibitem[{{Sorathia} {et~al.}(2012){Sorathia}, {Reynolds}, {Stone}, \&
  {Beckwith}}]{Sorathiaetal2012}
{Sorathia}, K.~A., {Reynolds}, C.~S., {Stone}, J.~M., \& {Beckwith}, K. 2012,
  \apj, 749, 189

\bibitem[{{Stella} \& {Rosner}(1984)}]{StellaRosner1984}
{Stella}, L., \& {Rosner}, R. 1984, \apj, 277, 312

\bibitem[{{Stone} {et~al.}(2008){Stone}, {Gardiner}, {Teuben}, {Hawley}, \&
  {Simon}}]{Stoneetal2008}
{Stone}, J.~M., {Gardiner}, T.~A., {Teuben}, P., {Hawley}, J.~F., \& {Simon},
  J.~B. 2008, \apjs, 178, 137

\bibitem[{{Tombesi} {et~al.}(2010){Tombesi}, {Cappi}, {Reeves}, {Palumbo},
  {Yaqoob}, {Braito}, \& {Dadina}}]{Tombesietal2010}
{Tombesi}, F., {Cappi}, M., {Reeves}, J.~N., {et~al.} 2010, \aap, 521, A57

\bibitem[{{Tombesi} {et~al.}(2015){Tombesi}, {Mel{\'e}ndez}, {Veilleux},
  {Reeves}, {Gonz{\'a}lez-Alfonso}, \& {Reynolds}}]{Tombesietal2015}
{Tombesi}, F., {Mel{\'e}ndez}, M., {Veilleux}, S., {et~al.} 2015, \nat, 519,
  436

\bibitem[{{Turner}(2004)}]{Turner2004}
{Turner}, N.~J. 2004, \apjl, 605, L45

\bibitem[{{Turner} {et~al.}(2003){Turner}, {Stone}, {Krolik}, \&
  {Sano}}]{Turneretal2003}
{Turner}, N.~J., {Stone}, J.~M., {Krolik}, J.~H., \& {Sano}, T. 2003, \apj,
  593, 992

\bibitem[{{Zheng} {et~al.}(1997){Zheng}, {Kriss}, {Telfer}, {Grimes}, \&
  {Davidsen}}]{Zhengetal1997}
{Zheng}, W., {Kriss}, G.~A., {Telfer}, R.~C., {Grimes}, J.~P., \& {Davidsen},
  A.~F. 1997, \apj, 475, 469

\end{thebibliography}
\end{CJK*}

\end{document}